\documentclass[11pt]{article}
\usepackage[margin=2cm,left=3cm,right=3cm]{geometry}

\usepackage{amsmath,amsthm,amsfonts,dsfont}
\usepackage{xcolor}
\usepackage{graphicx}
\usepackage{nicefrac}
\usepackage{xspace}
\usepackage{booktabs}
\usepackage{caption}
\usepackage{subcaption}
\usepackage{aligned-overset}
\usepackage{enumitem}
\usepackage{cite}
\usepackage{tikz}
\usetikzlibrary{automata, positioning}
\usepackage[lined,boxed,commentsnumbered]{algorithm2e}
\usepackage{makecell} 

\SetAlCapSkip{8pt}

\newtheorem{theorem}{Theorem}
\numberwithin{theorem}{section}
\newtheorem{lemma}[theorem]{Lemma}
\newtheorem{definition}[theorem]{Definition}

\newcommand{\eps}{\varepsilon}
\newcommand{\NN}{\mathds{N}}
\newcommand{\QQ}{\mathds{Q}}
\newcommand{\opt}{\operatorname{OPT}}

\renewcommand{\Pr}[1]{\mbox{\rm\bf Pr}\left[#1\right]}
\newcommand{\E}[1]{\mbox{\rm\bf E}\left[#1\right]}

\DeclareMathOperator{\pos}{pos}
\newcommand{\intD}[1]{~\mathrm{d} {#1}~}

\newtheorem{fact}{Fact}
\newtheorem{subfact}{Fact}[fact]

\newtheorem{observation}{Observation}

\title{Improved Online Algorithms for Knapsack and GAP \\ in the Random Order Model
	\thanks{Work supported by the European Research Council, Grant Agreement No.\ 691672. \\
		A preliminary version of this paper appeared in \textit{22nd International Conference on Approximation Algorithms for Combinatorial Optimization Problems (APPROX 2019).}}
}

\author{Susanne Albers \\
	Department of Computer Science, \\Technial University of Munich,\\
	albers@in.tum.de \and 
	Arindam Khan\\
	Department of Computer Science and Automation, \\Indian Institute of Science,\\
	arindamkhan@iisc.ac.in \and 
	Leon Ladewig\\
	Department of Computer Science, \\Technial University of Munich,\\
	ladewig@in.tum.de
}

\date{}

\begin{document}

\maketitle
\allowdisplaybreaks

\begin{abstract}
The {\em knapsack problem} is one of the classical problems in combinatorial optimization:
Given a set of items, each specified by its size and profit, the goal is to find a maximum profit packing into a knapsack of bounded capacity.
In the online setting, items are revealed one by one and the decision, if the current item is packed or discarded forever, must be done immediately and irrevocably upon arrival. 
We study the online variant in the random order model where the input sequence is a uniform random permutation of the item set.

We develop a randomized $(1/6.65)$-competitive algorithm for this problem, 
outperforming the current best algorithm of competitive ratio $1/8.06$ [Kesselheim et al.\ SIAM J. Comp. 47(5)].
Our algorithm is based on two new insights: We introduce a novel algorithmic approach that 
employs two given algorithms, optimized for restricted item classes, sequentially on the input sequence.
In addition, we study and exploit the relationship of the knapsack problem to the 2-secretary problem.

The {\em generalized assignment problem} (GAP) includes, besides the knapsack problem, several important problems related to scheduling and matching.
We show that in the same online setting, applying the proposed sequential approach yields a $(1/6.99)$-competitive randomized algorithm for GAP. Again, our proposed algorithm outperforms the current best result of competitive ratio $1/8.06$ [Kesselheim et al.\ SIAM J. Comp. 47(5)].

\end{abstract}

\section{Introduction}
\label{sec:intro}
Many real-world problems can be considered resource allocation problems.
For example, consider the loading of a cargo plane with (potential) goods of different weights.
Each item raises a certain profit for the airline if it is transported; however, not all goods can be loaded
due to airplane weight restrictions. 
Clearly, the dispatcher seeks for a maximum profit packing fulfilling the capacity constraint.
This example from \cite{DBLP:books/daglib/0010031} illustrates the \textit{knapsack problem}:
Given a set of $n$ items, specified
by a size and a profit value, and a resource (called knapsack) of fixed capacity, the goal
is to find a subset of items (called packing) with maximum total profit and whose total size does not exceed
the capacity.
Besides being a fundamental and extensively studied problem in combinatorial optimization,
knapsack problems arise in many and various practical settings.
We refer the readers to textbooks \cite{DBLP:books/daglib/0010031,Martello:1990:KPA:98124}
and to the surveys of previous work in \cite{ChristensenKPT17,GalvezGHI0W17} for further references.

The introductory example from cargo logistics can be generalized naturally to multiple airplanes of different capacities.
Here, the size and the profit of an item may depend on the airplane and on the schedule, respectively.
This leads to the \textit{generalized assignment problem} (GAP) \cite{Martello:1990:KPA:98124},
where resources of different capacities are given, and the size and the profit of an item depend on the 
resource to which it is assigned.
The GAP includes many prominent problems, such as the (multiple) knapsack problem \cite{DBLP:journals/siamcomp/ChekuriK05},
weighted bipartite matching \cite{DBLP:journals/tcs/KhullerMV94}, AdWords \cite{DBLP:journals/jacm/MehtaSVV07}, 
and the display ads problem \cite{DBLP:conf/wine/FeldmanKMMP09}.
Further applications of GAP are outlined in the survey articles \cite{CATTRYSSE1992260,DBLP:journals/infor/Oncan07}.

We study online variants of the knapsack problem and GAP. Here, $n$ items are presented sequentially, and the 
decision for each item must be made immediately upon arrival. 
This setting would arise in our logistics example if the dispatcher needs to answer customer requests immediately without knowledge of future requests.
In fact, many real-world optimization problems occur as online problems, as often decisions must be made under uncertain conditions.
The online knapsack problem has been studied in particular in the context of online auctions \cite{DBLP:conf/www/BorgsCIJEM07,DBLP:conf/wine/ZhouCL08}.

Typically, the performance measure for online algorithms is the \textit{competitive ratio}, which is defined as the ratio between the values of the algorithmic solution and an optimal offline solution for a worst-case input.
The knapsack problem admits no randomized algorithm of bounded competitive ratio in the general online setting  \cite{DBLP:conf/wine/ZhouCL08}.
This holds even if only a single item can be packed, as known from the secretary problem \cite{dynkin1963optimum,lindley1961dynamic}.
However, these hardness results are based on a worst-case input presented in adversarial order.
In the \textit{random order model}, the performance of an algorithm is evaluated for a worst-case input,
but the adversary has no control over the input order;
the input sequence is drawn uniformly at random among all permutations.

In order to define the competitive ratio of an algorithm $\mathcal{A}$ in this model formally, let $\mathcal{A}(\mathcal{I})$ and $\opt(\mathcal{I})$ denote the profits of the solutions of $\mathcal{A}$ and an optimal offline algorithm, respectively, for input $\mathcal{I}$.
We say that $\mathcal{A}$ is \textit{$r$-competitive} (or has \textit{competitive ratio} $r$) in the random order model if
\[
\E{\mathcal{A}(\mathcal{I})} \geq (r - o(1)) \cdot \opt(\mathcal{I})
\]
holds for all inputs $\mathcal{I}$. Here, the expectation is over the random permutation as well as over random choices of the algorithm. The $o(1)$-term is asymptotic with respect to the number $n$ of items in the input.

The random order model became increasingly popular in the field of online algorithms.
An early and well-known example is the secretary problem \cite{dynkin1963optimum,lindley1961dynamic}.
Nowadays, the matroid secretary problem \cite{DBLP:journals/jacm/BabaioffIKK18,DBLP:journals/mor/FeldmanSZ18} is considered as one of the most central problems in this field.
Further multiple-choice generalizations \cite{DBLP:conf/soda/ChanCJ15,DBLP:conf/soda/Kleinberg05} are part of active research as well.
The model has also been successfully applied to other problem classes including
scheduling \cite{DBLP:conf/icalp/AlbersJ20,DBLP:conf/esa/0002KT15,DBLP:conf/soda/Molinaro17},
packing \cite{DBLP:journals/ior/AgrawalWY14,DBLP:conf/esa/FeldmanHKMS10,DBLP:conf/soda/Kenyon96,DBLP:journals/siamcomp/KesselheimRTV18,DBLP:journals/mor/MolinaroR14},
graph problems \cite{DBLP:conf/soda/BahmaniMM10,DBLP:conf/esa/KesselheimRTV13,DBLP:conf/stoc/MahdianY11},
facility location \cite{DBLP:conf/focs/Meyerson01},
budgeted allocation \cite{DBLP:conf/soda/MirrokniGZ12}, and
submodular welfare maximization \cite{DBLP:journals/siamcomp/KorulaMZ18}.

\subsection{Related Work}

\paragraph{Online knapsack problem.}
The online knapsack problem was first studied by Marchetti-Spaccamela and Vercellis 
\cite{DBLP:journals/mp/Marchetti-SpaccamelaV95}, who showed that no deterministic online algorithm 
for this problem can obtain a constant competitive ratio.
Moreover, Chakrabarty et al.\ \cite{DBLP:conf/wine/ZhouCL08} demonstrated that this fact cannot be overcome by randomization.

Given such hardness results, several relaxations have been introduced and investigated.
Most relevant to our work are results in the random order model.
Introduced as the \textit{secretary knapsack problem} \cite{DBLP:conf/approx/BabaioffIKK07},
Babaioff et al.\ developed a randomized algorithm of competitive ratio
$1/(10e) < 1/27$.
Kesselheim et al.\ \cite{DBLP:journals/siamcomp/KesselheimRTV18} achieved a significant improvement by developing a $(1/8.06)$-competitive randomized algorithm for the generalized assignment problem.
Finally, Vaze \cite{DBLP:conf/infocom/Vaze17} showed that there exists a deterministic algorithm of competitive ratio $1/(2e) < 1/5.44$,
assuming that the maximum profit of a single item is small compared to the profit of the optimal solution.

Apart from the random order model, different further relaxations have been considered. Marchetti-Spaccamela and Vercellis \cite{DBLP:journals/mp/Marchetti-SpaccamelaV95} studied a stochastic 
model wherein item sizes and profits are drawn from a fixed distribution. 
Lueker \cite{DBLP:journals/jal/Lueker98} obtained improved bounds in this model.
Chakrabarty et al.\ \cite{DBLP:conf/wine/ZhouCL08} studied the problem 
when the density (profit-size ratio) of each item is in a fixed range $[L,U]$.
Under the further assumption that item sizes are small compared to the knapsack capacity,
Chakrabarty et al.\ proposed an algorithm of competitive ratio $\ln (U/L) + 1$ and provided a lower bound of
$\ln (U/L)$.
Another branch of research considers removable models, where the algorithm can
remove previously packed items. Removing such items can incur no cost
\cite{DBLP:journals/tcs/HanKM15,DBLP:conf/icalp/IwamaT02}
or a cancellation cost (\textit{buyback model}, \cite{babaioff2008selling,DBLP:conf/sigecom/BabaioffHK09,DBLP:journals/algorithmica/HanKM14}).
Recently, Vaze \cite{DBLP:conf/infocom/Vaze18} considered the problem under a (weaker) expected capacity constraint. This variant admits a competitive ratio of $1/4e$.

\paragraph{Online GAP.}
Since all hardness results for online knapsack also hold for online GAP,
research focuses on stochastic variants or modified online settings.
Currently, the only result for the random order model is the previously mentioned
$(1/8.06)$-competitive randomized algorithm proposed by Kesselheim et al.\ \cite{DBLP:journals/siamcomp/KesselheimRTV18}.
To the best of our knowledge, the earliest paper considering online GAP is due to Feldman et al.\ \cite{DBLP:conf/wine/FeldmanKMMP09}. 
They obtained an algorithm of competitive ratio tending to $1-1/e$ in the \textit{free disposal model}. In this model, the total size of items assigned to a resource might exceed its capacity; in addition, no item consumes more than a small fraction of any resource.
A stochastic variant of online GAP was studied by Alaei et al.\ \cite{DBLP:conf/approx/AlaeiHL13}.
Here, the size of an item is drawn from an individual distribution that is revealed upon arrival of the item, together with its profit. However, the algorithm learns the actual item size only after the assignment.
If no item consumes more than a $(1/k)$-fraction of any resource, the algorithm proposed by Alaei et al.\ has competitive ratio $1 - 1/\sqrt{k}$.

\paragraph{Online packing LPs.} Packing problems where requests can consume $d \geq 1$ different resources lead to general online packing LPs. Note that the special case of $d=1$ is the generalized assignment problem.
Buchbinder and Naor \cite{DBLP:journals/mor/BuchbinderN09} initiated the study of online packing LPs in the adversarial model.
The random order model admits $(1-\eps)$-competitive algorithms assuming large capacity ratios, i.e., the capacity of any resource is large compared to the maximum demand for it. This has been shown in a sequence of papers \cite{DBLP:journals/ior/AgrawalWY14,DBLP:conf/esa/FeldmanHKMS10,DBLP:journals/siamcomp/KesselheimRTV18,DBLP:journals/mor/MolinaroR14}. 
Recently, Kesselheim et al.\ \cite{DBLP:journals/siamcomp/KesselheimRTV18} gave an algorithm of competitive ratio $1 - O(\sqrt{(\log d) / B})$ where $B$ is the capacity ratio. Consequently, their algorithm is $(1-\eps)$-competitive if $B = \Omega((\log d) / \eps^2)$. For $d=1$, this result matches the lower bound by Kleinberg \cite{DBLP:conf/soda/Kleinberg05}.

\subsection{Our Contributions}
\label{sec:contribution}
As outlined above, for online knapsack and GAP in the adversarial input model, nearly
all previous works attain constant competitive ratios at the cost of either (a) imposing structural
constraints on the input or (b) significantly relaxing the original online model. 
Therefore, we study both problems in the random order model, which is less pessimistic than the
adversarial model but still considers worst-case instances without further constraints on the item properties.
For the knapsack problem, our main result is the following.

\begin{theorem}
	\label{thm:KSmainTheorem}	
	There exists a $(1/6.65)$-competitive randomized algorithm for 
	the online knapsack problem in the random order model.
\end{theorem}
One challenge in the design of knapsack algorithms is that the optimal packing can have, on a high level,  at least two different structures. Either there are a few large items, constituting the majority of the packing's profit, or there are many small such items.
Previous work \cite{DBLP:conf/approx/BabaioffIKK07,DBLP:journals/siamcomp/KesselheimRTV18} is based on splitting the input according to item sizes and then employing algorithms tailored for these restricted instances. However, the algorithms from  \cite{DBLP:conf/approx/BabaioffIKK07,DBLP:journals/siamcomp/KesselheimRTV18} choose a single item type via an initial random choice,
and then pack items of that type exclusively.
In contrast, our approach considers different item types in distinct time intervals, rather than discarding items of a specific type in advance.
More precisely, we develop algorithms $\mathcal{A}_L$ and $\mathcal{A}_S$ which are combined in a novel 
\textit{sequential approach}: 
While large items appearing in early rounds are packed using $\mathcal{A}_L$, algorithm
$\mathcal{A}_S$ is applied to pack small items revealed in later rounds.
We think that this approach may be helpful for other problems in similar online settings as well.

The proposed algorithm $\mathcal{A}_L$ deals with the knapsack problem where all items consume more than 
$1/3$ of the capacity (we call this problem 2-KS).
The 2-KS problem is closely related to the $k$-secretary problem \cite{DBLP:conf/soda/Kleinberg05} for $k=2$.
We also develop a general framework that allows to employ any algorithm for the 2-secretary problem to obtain an algorithm for 2-KS. 
As a side product, we obtain a simple $(1/3.08)$-competitive deterministic algorithm for 2-KS in the random order model.
For items whose size is at most $1/3$ of the resource capacity,
we give a simple and efficient algorithm $\mathcal{A}_S$.
Here, a challenging constraint is that $\mathcal{A}_L$ and $\mathcal{A}_S$ share the same resource, 
so we need to argue carefully that the decisions of $\mathcal{A}_S$ are feasible, given the
packing of $\mathcal{A}_L$ from previous rounds.

Finally, we show that the proposed sequential approach also improves the current best result for GAP 
\cite{DBLP:journals/siamcomp/KesselheimRTV18} from competitive ratio $1/8.06$ to $1/6.99$.
\begin{theorem}
	\label{thm:GAPmainTheorem}		
	There exists a $(1/6.99)$-competitive randomized algorithm for 
	the online generalized assignment problem in the random order model.
\end{theorem}
For this problem, we use the algorithmic building blocks $\mathcal{A}_L$, $\mathcal{A}_S$
developed in \cite{DBLP:conf/esa/KesselheimRTV13,DBLP:journals/siamcomp/KesselheimRTV18}.
However, we need to verify that $\mathcal{A}_L$, an algorithm for edge-weighted bipartite matching \cite{DBLP:conf/esa/KesselheimRTV13}, satisfies the desired properties for the sequential approach.
We point out that the assignments of our algorithm differ structurally from the assignments of the algorithm proposed in \cite{DBLP:journals/siamcomp/KesselheimRTV18}.
In the assignments of the latter algorithm, all items are either large or small compared to the capacity
of the assigned resource. 
In our approach, both situations can occur, because resources are managed independently.

\paragraph{Roadmap.}
We focus on the result on the knapsack problem (Theorem~\ref{thm:KSmainTheorem}) in the first sections of this paper. For this purpose, we provide elementary definitions and facts in Section~\ref{sec:preliminaries}.
Our main technical contribution is formally introduced in Section~\ref{sec:sequentialApproach}:
Here, we describe an algorithmic framework performing two algorithms $\mathcal{A}_L$, $\mathcal{A}_S$ sequentially.
In Sections \ref{sec:largeItems} and \ref{sec:smallItems}, we design and analyze the algorithms $\mathcal{A}_L$ and $\mathcal{A}_S$ for the knapsack problem.
Finally, in Section~\ref{sec:gap} we describe how the sequential approach can be applied to GAP.

\section{Preliminaries}
\label{sec:preliminaries}

Let $[n] := \{1,\ldots,n\}$. Further, let $\QQ_{\geq 0}$ and $\QQ_{>0}$ denote the set of non-negative 
and positive rational numbers, respectively.

\paragraph{Knapsack problem.}
We are given a set of items $ I = [n]$, 
each item $i \in I$ has \textit{size} $s_i \in \QQ_{> 0}$ and a \textit{profit} (\textit{value}) 
$v_i \in \QQ_{\geq 0}$.
The goal is to find a maximum profit packing into a knapsack of size $W \in \QQ_{> 0}$, i.e., a subset 
$M \subseteq I$ such that $\sum_{i \in M} s_i \leq W$ and $\sum_{i \in M} v_i$ is maximized.
W.l.o.g.\ we can assume $s_i \leq W$ for all $i \in I$.
In the online variant of the problem, a single item $i$ is revealed together with its size and profit in each \textit{round} $\ell \in [n]$. The online algorithm must decide immediately and irrevocably whether to pack $i$. We call an item \textit{visible in round $\ell$} if it arrived in round $\ell$ or earlier.

We classify items as large or small, depending on their size compared to $W$ and a parameter $\delta \in (0,1)$ to be determined later.

\begin{definition}
	\label{def:itemSizes}
	We say an item $i$ is $\delta$-\textit{large} if $s_i > \delta W$ and 
	$\delta$-\textit{small} if $s_i \leq \delta W$. 
	Whenever $\delta$ is clear from the context, we say an item is \textit{large}
	or \textit{small} for short. Based on the given item set $I$, we define two modified item sets $I_L$ and $I_S$, which are obtained as follows:
	\begin{itemize}
		\setlength\itemsep{0em}	
		\item $I_L$: Replace each small item by a large item of profit 0
		\item $I_S$: Replace each large item by a small item of profit 0.	
	\end{itemize}
\end{definition}
\noindent
Therefore, $I_L$ only contains large items and $I_S$ only contains small items.
We can assume that no algorithm packs a zero-profit item, thus any algorithmic packing of $I_L$ or $I_S$ can be turned into a packing of $I$ having the same profit.
Let $\opt$, $\opt_L$, and $\opt_S$ be the total profits of optimal packings for $I$, $I_L$, and $I_S$, respectively.
A useful upper bound for $\opt$ is 
\begin{equation}
\label{eq:optL+optS}
\opt \leq \opt_L + \opt_S.
\end{equation}

\paragraph{Bounding sums by integrals.}
In order to obtain lower or upper bounds on sums in closed form, we often make use of the following facts.

\setcounter{fact}{1}
\begin{subfact}
\label{fact:approxSumByIntegralDec}	
	Let $f$ be a non-negative real-valued function and let $a,b \in \NN$. If $f$ is monotonically decreasing, then 
	$\int_{a}^{b+1} f(i) \intD{i} \leq \sum_{i=a}^{b} f(i) \leq \int_{a-1}^{b} f(i) \intD{i}$.
\end{subfact}

\begin{subfact}
\label{fact:approxSumByIntegralInc}	
	Let $f$ be a non-negative real-valued function and let $a,b \in \NN$. If $f$ is monotonically increasing, then 
	$\int_{a-1}^{b} f(i) \intD{i} \leq \sum_{i=a}^{b} f(i) \leq \int_{a}^{b+1} f(i) \intD{i}$.
\end{subfact}

\section{Sequential Approach}
\label{sec:sequentialApproach}

\begin{algorithm}[t]
	\SetKwInOut{Input}{Input}
	\SetKwInOut{Output}{Output}
	\Input{Random permutation $\pi$ of $n$ items in $I$, a knapsack of capacity $W$,  \\
		parameters $c,d \in (0,1)$ with $c<d$, 
		algorithms $\mathcal{A}_L$, $\mathcal{A}_S$.}
	\Output{A feasible (integral) knapsack packing.}
	
	Let $\ell$ be the current round. \\
	\uIf{$\ell \leq cn$}{Sampling phase -- discard all items;}
	\uIf{$cn+1 \leq \ell \leq dn$}{Pack $\pi(\ell)$ iff $\mathcal{A}_L$ packs $\pi_L(\ell)$;}
	\uIf{$dn+1 \leq \ell \leq n$}{Pack $\pi(\ell)$ iff $\mathcal{A}_S$ packs $\pi_S(\ell)$ 
		and the remaining capacity is sufficiently large.}
	
	\caption{Sequential approach}
	\label{alg:sequentialAlg}
\end{algorithm}

A common approach in the design of algorithms for secretary problems is to set two phases: a \textit{sampling phase}, where all items are rejected, followed by a \textit{decision phase}, where some items are accepted according to a decision rule. Typically, this rule is based on the information gathered in the sampling phase.
We take this concept a step further: The key idea of our sequential approach is to use a part of the sampling phase of one algorithm as decision phase of another algorithm, which itself can have a sampling phase. This way, two algorithms are performed in a sequential way, which makes better use of the entire instance. 
We combine this idea with using different strategies for small and large items.

Formally, let $\mathcal{A}_L$ and $\mathcal{A}_S$ be two online knapsack algorithms
and $I_L$ and $I_S$ be the item sets constructed according to Definition~\ref{def:itemSizes}.
Further, let $0<c<d<1$ be two parameters to be specified later.
Our proposed algorithm samples the first $cn$ rounds; no item is packed during this time.
From round $cn+1$ to $dn$, the algorithm considers large items exclusively.
In this interval it follows the decisions of $\mathcal{A}_L$. 
After round $dn$, the algorithm processes only small items and follows the decisions of $\mathcal{A}_S$. However, it might be the case that an item accepted by $\mathcal{A}_S$ cannot be packed because the knapsack capacity is exhausted due to the packing of $\mathcal{A}_L$ in earlier rounds.
Note that all rounds $1,\ldots,dn$ can be considered as the sampling phase for $\mathcal{A}_S$.
A formal description is given in Algorithm~\ref{alg:sequentialAlg}.
Here, for a given input sequence $\pi$ of $I$, let $\pi_L$ and $\pi_S$ denote the corresponding sequences
from $I_L$ and $I_S$, respectively. Note that $\pi$ is revealed sequentially and 
$\pi_L$, $\pi_S$ can be constructed online. 
For any input sequence $\pi$, let $\pi(\ell)$ denote the item at position $\ell \in [n]$.

In the final algorithm, we set the threshold for small items to $\delta=1/3$ and use
Algorithm~\ref{alg:sequentialAlg} with parameters $c=0.42291$ and $d=0.64570$.
The choice of $c$ and $d$ maximizes the minimum of 
$\E{\mathcal{A}_L} / \opt_L$ and $\E{\mathcal{A}_S} / \opt_S$.
For simplicity, we assume $cn,dn \in \NN$. If $n$ is large enough, this assumption does not affect the competitive ratio substantially.
We next give a high-level description of the proof of Theorem~\ref{thm:KSmainTheorem}.

\begin{proof}[Proof of Theorem~\ref{thm:KSmainTheorem}]
	Let $\mathcal{A}$ be Algorithm~\ref{alg:sequentialAlg} and let $\mathcal{A}_L$, $\mathcal{A}_S$ be the
	algorithms developed in Sections~\ref{sec:largeItems} and \ref{sec:smallItems}.
	In the next sections, we prove the following results for $r= 1/6.65 - o(1)$
	(see Lemmas \ref{lemma:2secFinalParameters} and \ref{lemma:KSsmallItemsFinal}):
	The expected profit from $\mathcal{A}_L$ in rounds $cn+1,\ldots,dn$ is at least $r \cdot \opt_L$,
	and the expected profit from $\mathcal{A}_S$ in rounds $dn+1,\ldots,n$ is at least $r \cdot \opt_S$.
	Together with inequality~(\ref{eq:optL+optS}), we obtain
	\begin{equation*}
	\E{\mathcal{A}}
	\geq \E{\mathcal{A}_L} + \E{\mathcal{A}_S} 
	\geq r \cdot \opt_L + r \cdot \opt_S
	\geq \left( \frac{1}{6.65} - o(1) \right) \opt \,.
	\qedhere
	\end{equation*}
\end{proof}

The order in which $\mathcal{A}_L$ and $\mathcal{A}_S$ are arranged in Algorithm~\ref{alg:sequentialAlg} follows from two observations. Algorithm $\mathcal{A}_S$ is powerful if it samples roughly $2n/3$ rounds;
a part of this long sampling phase can be used as the decision phase of $\mathcal{A}_L$, for which a shorter sampling phase is sufficient. 
Moreover, the first algorithm should either pack high-profit items, or should leave the knapsack empty for the following algorithm with high probability.
The algorithm $\mathcal{A}_L$ we propose in Section~\ref{sec:largeItems} has this property (see Lemma~\ref{lemma:2SecEmptyKnapsack}), in contrast to $\mathcal{A}_S$. If $\mathcal{A}_S$ would precede $\mathcal{A}_L$, the knapsack would be empty after round $dn$ with very small probability, in which case we would not benefit from $\mathcal{A}_L$ at all.

Finally, note that stronger algorithms for the respective sub-problems can be obtained by choosing different parameters or algorithmic approaches (see Lemma~\ref{lemma:2secD1} and \cite{DBLP:journals/siamcomp/KesselheimRTV18}). 
However, we seek for maximizing the competitive ratio of Algorithm~\ref{alg:sequentialAlg} and therefore need algorithms $\mathcal{A}_L$ and $\mathcal{A}_S$ that perform well within the sequential framework.

\section{Large Items}
\label{sec:largeItems}

\begin{algorithm}[t]
	\SetKwInOut{Input}{Input}
	\SetKwInOut{Output}{Output}
	\Input{Random permutation of $n$ $(1/3)$-large items, a knapsack of capacity $W$, \\
		parameters $c,d \in (0,1)$ with $c<d$.}
	\Output{A feasible (integral) packing of the knapsack.}
	
	Let $\ell$ be the current round. \\
	\uIf{$\ell \leq cn$}{Sampling phase -- discard all items.}
	Let $v^*$ be the maximum profit seen up to round $cn$. \\
	\uIf{$cn+1 \leq \ell  \leq dn$}{Pack the first two items of profit higher than $v^*$, if feasible. \\
	}
	\uIf{$\ell > dn$}{Discard all items.
	}	
	\caption{Algorithm $\mathcal{A}_L$ for large items}
	\label{alg:2sec}
\end{algorithm}

The approach presented in this section is based on the connection between the online knapsack problem under random arrival order and the $k$-secretary problem \cite{DBLP:conf/soda/Kleinberg05}.
In the latter problem, the algorithm can accept up to $k$ items and the goal is to maximize the sum of their profits. 
Therefore, we assume that a $k$-secretary algorithm can observe the actual profits of the items, as opposed to the ordinal version of the problem, where an algorithm can only decide based on relative merits.
This way, the $k$-secretary problem generalizes the classical secretary problem \cite{dynkin1963optimum,lindley1961dynamic}
and is itself a special case of the online knapsack problem under random arrival order (if all knapsack items have size $W/k$).

In our setting, each large item consumes more than $\delta=1/3$ of the knapsack capacity. We call this problem 2-KS, since at most two items can be packed completely. Therefore, any $2$-secretary algorithm can be employed to identify two high-profit items for the knapsack packing. However, after packing the first item, the resource might be exhausted, such that the second item identified by the 2-secretary algorithm cannot be packed.

Although this idea can be generalized to any $k$-secretary algorithm and corresponding $\delta$-large items, the approach seems stronger for small $k$: While $1$-KS is exactly $1$-secretary, the characteristics of $k$-KS and $k$-secretary deviate with growing $k$. Our results show that the problems $2$-secretary and $2$-knapsack are still close enough to benefit from such an approach.

In the following, let $\mathcal{A}_L$ be Algorithm~\ref{alg:2sec}.
This is an adaptation of the algorithm \textsc{single-ref} developed for the $k$-secretary problem in \cite{kSecretary}. As discussed above, 2-secretary and 2-KS are similar, but different problems.
Therefore, in our setting it is not possible to apply the existing analysis from \cite{kSecretary} or from any other $k$-secretary algorithm directly.
We further note that in the approach described below, in principle any 2-secretary algorithm can be employed.
In Section~\ref{sec:kSecAlgorithms}, we discuss several alternative algorithms.

\paragraph{Assumption.}
For this section, we assume that all profits are distinct. This is without loss of generality, as ties can be broken by adjusting the profits slightly, using the items' identifiers. Further, we assume $v_1 > v_2 > \ldots > v_n$
and say that $i$ is the \textit{rank} of item $i$.

\subsection{Packing Types}
\label{sec:ksPackingTypes}

As outlined above, in contrast to the 2-secretary problem, not all combinations of two knapsack items can be packed completely.
Therefore, we analyze the probability that $\mathcal{A}_L$ selects a feasible set of items 
whose profit can be bounded from below.
We restrict our analysis to packings where an item $i \in \{1,2,3,4\}$ is packed as the first item
and group such packings into several packing types A-M defined in the following.
Although covering more packings might lead to further insights into the problem and to a stronger result, we expect the improvement to be marginal. 

Let $p_X$ be the probability that $\mathcal{A}_L$ returns a packing of type $X \in \{\text{A},\ldots,\allowbreak \text{M}\}$. 
In addition, let $p_i$ for $i \in [n]$ be the probability that $\mathcal{A}_L$ packs $i$ as the first item.
Finally, let $p_{ij}$ for $i,j \in [n]$ be the probability that $\mathcal{A}_L$ packs $i$ as the first item and $j$ as the second item.

\begin{table}
	\centering
	\caption{Definition of packing types A-M. 
		We use set notation $\{i,j\}$ if $i$ and $j$ can be packed in any order, and tuple notation $(i,j)$ if the packing order must be as given. \strut}
	\label{tab:packingTypes}
	\begin{tabular}{lllll} \toprule
		Type & Content & Constraint on $j$ & Probability $p_X$ \\ \midrule
		A & $\{1,2 \}$  & -  & $p_{12} + p_{21}$ \\ 
		B & $\{1,3\}$   & -  &$p_{13} + p_{31}$  \\			
		C & $\{2,3 \}$   & -  &$p_{23} + p_{32}$ \\ \midrule
		D & $(1,j)$         & - & $p_1$ \\
		E & $(2,j)$   		& - &  $p_2$\\  
		F & $(3,j)$   		& - & $p_3$ \\  
		G & $(4,j)$   		& - & $p_4$ \\  \midrule			
		H & $(1,j)$  	   & $j \neq 2$ & $p_1 - p_{12}$            \\
		I & $(1,j)$          & $j \neq 3$ & $p_1 - p_{13}$ \\ \midrule
		J & $(2,j)$          & $j \neq 1$  & $p_2 - p_{21}$\\
		K & $(2,j)$   		& $j \neq 3$ & $p_2 - p_{23}$ \\  \midrule
		L & $(3,j)$   		& $j \neq 1$ & $p_3 - p_{31}$ \\  
		M & $(3,j)$   		& $j \neq 2$ & $p_3 - p_{32}$ \\  \bottomrule			
	\end{tabular} 
\end{table}
In a packing of type A, the items 1 and 2 are packed in any order. Therefore, $p_\text{A} = p_{12} + p_{21}$. 
The types B and C are defined analogously using the items $\{1,3\}$ and $\{2,3\}$, respectively.
In a packing of type D, the item $1$ is accepted as the first item, together with no or any second item $j$. This happens with probability $p_\text{D} = p_1$. Accordingly, we define types E, F, and G using the items 2, 3, and 4, respectively.
Finally, for each item $i \in \{1,2,3\}$, we introduce two further packing types. For $i=1$, types H and I cover packings where the first accepted item is 1, the second accepted item $j$ is not 2 (type H) and not 3 (type I), respectively. Therefore, we get $p_\text{H} = p_1 - p_{12}$ and $p_\text{I} = p_1 - p_{13}$. Packing types J-K and L-M describe analogous packings for $i=2$ and $i=3$, respectively.
Table~\ref{tab:packingTypes} shows all packing types A-M and their probabilities expressed by $p_i$ and $p_{ij}$.

In Section~\ref{sec:2secCases}, we use the packing types to describe a subset of packings whose profit can be bounded against $\opt_L$.
For example, suppose that $\opt_L = v_1 + v_2$. Then, all relevant packings are of type A, H, or J. As these types are disjoint by definition, we immediately obtain
$\E{\mathcal{A}_L} \geq p_\text{A} (v_1 + v_2) + p_\text{H} v_1 + p_\text{J} v_2$.
\subsection{Acceptance Probabilities of Algorithm~\ref{alg:2sec}}

In the following, we compute the probabilities $p_i$ and $p_{ij}$ from Table~\ref{tab:packingTypes} as functions of $c$ and $d$.
Throughout the following proofs, we denote the position of an item $i$ in a given permutation with $\pos(i) \in [n]$.
Further, let $a$ be the maximum profit item from the sampling.

We think of the random permutation as being sequentially constructed. The fact given below follows from the hypergeometric distribution and becomes helpful in the proofs of Lemmas~\ref{lemma:2secSingleItemProbabilitiies} and \ref{lemma:2secPairProbabilities}.

\begin{fact}
	\label{fact:BlueBalls}
	Suppose there are $N$ balls in an urn from which $M$ are blue and $N-M$ red.
	The probability of drawing $K$ blue balls without replacement in a sequence of length $K$ is 
	$ h(N,M,K) := \binom{M}{K} / \binom{N}{K} \,.$
\end{fact}

In the first lemma, we provide the exact probability $p_i$ for all $i \in [n]$ and give lower bounds for $p_i$ when $i \in [4]$.

\begin{figure*}
	\centering
	\includegraphics[width=0.9\textwidth]{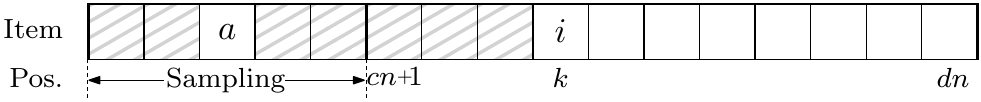}
	\caption{Input sequence considered in Lemma~\ref{lemma:2secSingleItemProbabilitiies}. The gray dashed slots represent items of rank greater than $a$.}
	\label{fig:eventE}
\end{figure*}
\begin{lemma}
	\label{lemma:2secSingleItemProbabilitiies}
	The probability that item $i \in [n]$ is accepted as the first item is
	\[
	p_i = \frac{c}{n-1} \sum_{k=cn+1}^{dn} \frac{\binom{n-i}{k-1}}{\binom{n-2}{k-2}} \,.
	\]	
	Moreover, we have the lower bound
	\[
	p_i \geq 
	\begin{cases}
	c \ln \frac{d}{c} - o(1) & i=1 \\
	c \left( \ln \frac{d}{c} - d + c \right) - o(1) & i=2 \\
	c \left( \ln \frac{d}{c} - 2(d-c) + \frac{1}{2} (d^2 - c^2) \right ) - o(1) & i=3 \\
	c \left( \ln \frac{d}{c} -3(d-c) + \frac{3}{2} (d^2-c^2) - \frac{1}{3} (d^3-c^3) \right ) - o(1) & i=4 \,.
	\end{cases} \,
	\]	
\end{lemma}

\begin{proof}
	In the first part of this proof, we analyze the probability that item $i$ is accepted as the first item at a fixed position $k \geq cn+1$.
	As $a$ is defined as the best sampling item, $\pos(a) \leq cn$ must hold.
	A permutation uniformly drawn at random satisfies $\pos(i)=k$ and $\pos(a) \leq cn$ with
	probability $\frac{1}{n} \frac{cn}{n-1} = \frac{c}{n-1}$.
	Next, we draw the remaining $k-2$ items for the positions before $k$ (see Figure~\ref{fig:eventE}).
	Since $i$ is packed as the first item, all previous items (except for $a$) must have rank greater than $a$. 
	As these items are drawn from the remaining $n-2$ items (of which $n-a$ have rank greater than $a$), the probability for this step is $h(n-2,n-a,k-2)$ according to Fact~\ref{fact:BlueBalls}. 
	Using the law of total probability for $k \in \{cn+1,\ldots,dn\}$ and $a \in \{i+1,\ldots,n\}$, we obtain
	\begin{align}
	p_i
	&= \frac{c}{n-1} \sum_{k=cn+1}^{dn} \sum_{a=i+1}^{n} h(n-2,n-a,k-2)  \notag \\
	&= \frac{c}{n-1} \sum_{k=cn+1}^{dn} \frac{1}{\binom{n-2}{k-2}} \sum_{a=i+1}^{n} \binom{n-a}{k-2} \notag \\
	&= \frac{c}{n-1} \sum_{k=cn+1}^{dn} \frac{\binom{n-i}{k-1}}{\binom{n-2}{k-2}} 	\label{eq:exactSingleItemProb} \,.
	\end{align}
	Here, the last identity follows from
	$
	\sum_{a=i+1}^{n} \binom{n-a}{k-2} = \sum_{a=0}^{n-i-1} \binom{a}{k-2} = \binom{n-i}{k-1} \,.
	$
	
	In the second part of the proof, we derive a lower bound for $p_i$. We first consider the quotient of binomial coefficients from Equation~(\ref{eq:exactSingleItemProb}) and observe
	\begin{align}
	\frac{\binom{n-i}{k-1}}{\binom{n-2}{k-2}}
	&= \frac{(n-i)!}{(k-1)! \cdot (n-i-k+1)!} \cdot \frac{(k-2)! \cdot (n-k)!}{(n-2)!} \notag \\
	&= \frac{(n-i)!}{(n-2)!} \cdot \frac{(n-k)!}{(n-i-k+1)!} \cdot \frac{1}{k-1} \notag \\
	&> \frac{1}{(n-2)^{i-2}} \cdot \frac{(n-k)!}{(n-i-k+1)!} \cdot \frac{1}{k} \notag \\
	&> \frac{(n-k-i)^{i-1}}{n^{i-2}} \cdot \frac{1}{k} \label{eq:BoundRatioOfBC} \,.
	\end{align}
	Combining Equation~(\ref{eq:exactSingleItemProb}) and inequality~(\ref{eq:BoundRatioOfBC}) yields
	\begin{equation}
	\label{ineq:piInterim}
	p_i > \frac{c}{n-1} \sum_{k=cn+1}^{dn} \frac{(n-k-i)^{i-1}}{n^{i-2}} \cdot \frac{1}{k}
	    > \frac{c}{n^{i-1}} \sum_{k=cn+1}^{dn} \frac{(n-k-i)^{i-1}}{k} \,.
	\end{equation}
	Now, the goal is to find a closed expression which bounds the last sum in inequality~(\ref{ineq:piInterim}) from below. We have
	\begin{equation}
	\label{ineq:piInterim2}
	\sum_{k=cn+1}^{dn} \frac{(n-k-i)^{i-1}}{k}
	= \sum_{k=cn+1+i}^{dn+i} \frac{(n-k)^{i-1}}{k-i}
	> \sum_{k=cn+1+i}^{dn+i} \frac{(n-k)^{i-1}}{k}
	\end{equation}
	and define $f(k) = (n-k)^{i-1} / k$. Since $f$ is monotonically decreasing in $k$ and $i-1 \geq 0$,
	we have
	\begin{multline}
	\label{ineq:piInterim3}
	\sum_{k=cn+1+i}^{dn+i} \frac{(n-k)^{i-1}}{k}
	= \sum_{k=cn}^{dn-1} f(k) + \sum_{k=dn}^{dn+i} f(k) - \sum_{k=cn}^{cn+i} f(k) \\
	> \int_{cn}^{dn} f(k) \intD{k} - (i+1) \cdot f(cn) 
	= \int_{cn}^{dn} f(k) \intD{k} - (i+1) \cdot \frac{(n-cn)^{i-1}}{cn} \,,
	\end{multline}
	where we used that Fact~\ref{fact:approxSumByIntegralDec}. 
	Let $F$ be a function such that $\int_{cn}^{dn} f(k) \intD{k} = F(dn)-F(cn)$. By combining inequalities
	(\ref{ineq:piInterim}) to (\ref{ineq:piInterim3}) we obtain
	\begin{equation}
	\label{ineq:singleItemIntegral}
	p_i > \frac{c}{n^{i-1}} \cdot \left(F(dn) - F(cn) \right) - (i+1) \cdot \frac{(1-c)^{i-1}}{n} \,. 
	\end{equation}
	Below we provide suitable functions $F$ for $i \in [4]$. \\[1em]
\begin{center}
		\begin{tabular}{llll}
		$i$ &  $f(k)$ &  $F(k)$ &  $F(dn) - F(cn)$ \\ \toprule[1pt] 
		$1$ &  $\frac{1}{k}$ &  $\ln k$ &  $\ln \frac{d}{c}$ \\ \midrule
		$2$ &  $\frac{n-k}{k}$ &  $n \ln k - k$ &  $n \ln \frac{d}{c} -dn +cn$ \\ 		\midrule
		$3$ &  $\frac{(n-k)^2}{k}$ &  $n^2 \ln k - 2nk + \frac{k^2}{2}$ 
		&  $n^2 \ln \frac{d}{c} -2n(dn-cn) + \frac{d^2 n^2 - c^2 n^2}{2}$ \\ 		\midrule
		$4$ &  $\frac{(n-k)^3}{k}$ &  \makecell[l]{$n^3 \ln k -3 n^2 k$\\[0.5em] $+ \frac{3}{2} nk^2 - \frac{k^3}{3}$} 
		& \makecell[l]{$	n^3 \ln \frac{d}{c} - 3n^3 (d-c)$\\[0.5em] $+ \frac{3}{2} n^3 (d^2-c^2) - \frac{1}{3} n^3(d^3-c^3) $} \\ 		
	\end{tabular} \\[1em]
\end{center}
	The claim follows by substituting $F(dn)-F(cn)$ in inequality~(\ref{ineq:singleItemIntegral}) by the corresponding expression from the table and noting that $(i+1) \cdot \frac{(1-c)^{i-1}}{n} = o(1)$.
\end{proof}

Next, we analyze the probabilities $p_{ij}$ with $i < j$ and give lower bounds for $p_{12}$, $p_{13}$, and $p_{23}$.
\begin{lemma}
	\label{lemma:2secPairProbabilities}
	Let $i$ and $j$ be two items with $i < j$. The probability that $i$ is selected as the first item and $j$ is selected as the second item is
	\[
	p_{ij} = \frac{c}{n-1} \cdot \frac{1}{n-2} \cdot \sum_{k=cn+1}^{dn-1} \sum_{l=k+1}^{dn} \frac{\binom{n-j}{l-2}}{\binom{n-3}{l-3}} \,.
	\]
	Moreover, it holds that
	\begin{align*}
		p_{12} 			&\geq c \left( d - c \ln \frac{d}{c} -c \right) - o(1) \,, \\
		p_{13} = p_{23} &\geq c \left( d - c \ln \frac{d}{c} -c - \frac{d^2}{2} +cd - \frac{c^2}{2} \right) - o(1) \,.
	\end{align*}
\end{lemma}
\noindent

\begin{figure}
	\centering
	\includegraphics[width=0.9\textwidth]{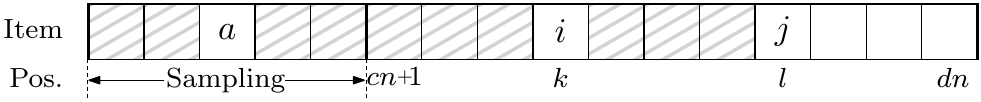}
	\caption{Input sequence considered in Lemma~\ref{lemma:2secPairProbabilities}. The gray dashed slots represent items of rank greater than $a$.}
	\label{fig:eventLemmaPair}
\end{figure}

\begin{proof}
	Let $i$, $j$ be two items with $i < j$.
	The proof follows the same structure as the proof of Lemma~\ref{lemma:2secSingleItemProbabilitiies}.
	Again, we construct the permutation by drawing the positions for items $i$, $j$, and $a$ first 
	and afterwards all remaining items with position up to $\pos(j)$ (see Figure~\ref{fig:eventLemmaPair}).
	Fix positions $k=\pos(i)$ and $l=\pos(j)$ .
	Again, $\pos(a) \leq cn$ must hold by definition of $a$. 
	The probability that a random permutation satisfies these three position constraints is
	$\beta := \frac{1}{n} \frac{1}{n-1} \frac{cn}{n-2} = \frac{c}{n-1} \cdot \frac{1}{n-2}$. 
	All remaining items up to position $l$ must have rank greater than $a$.
	Thus, we need to draw $l-3$ items from a set of $n-3$ remaining items, from which $n-a$ have rank greater than $a$.
	This happens with probability $h(n-3,n-a,l-3)$. 
	Using the law of total probability for $k$, $l$ with $cn+1 \leq k < l \leq dn$ and $a \in \{j+1,\ldots,n\}$, we obtain
	\begin{align*}
	p_{ij} 
	&= \beta \sum_{k=cn+1}^{dn-1} \sum_{l=k+1}^{dn} \sum_{a=j+1}^{n} h(n-3,n-a,l-3) \\
	&= \beta \sum_{k=cn+1}^{dn-1} \sum_{l=k+1}^{dn} 
	\frac{1}{\binom{n-3}{l-3}} \sum_{a=j+1}^{n} \binom{n-a}{l-3} \,.
	\end{align*}
	Again, by observing 
	$\sum_{a=j+1}^{n} \binom{n-a}{l-3} = \sum_{a=0}^{n-j-1} \binom{a}{l-3} = \binom{n-j}{l-2}$, we obtain finally
	\begin{equation}
	\label{eq:exactPairProbability}
	p_{ij} = \beta \sum_{k=cn+1}^{dn-1} \sum_{l=k+1}^{dn} \frac{\binom{n-j}{l-2}}{\binom{n-3}{l-3}} \,.
	\end{equation}
	
	To prove the second part of the lemma, first note that Equation~(\ref{eq:exactPairProbability}) does not depend on $i$, thus we have $p_{13} = p_{23}$.
	It remains to find lower bounds for $p_{12}$ and $p_{23}$.
	We start with $p_{12}$. By Equation~(\ref{eq:exactPairProbability}) and the definition of $\beta$, it holds that
	\begin{equation}
	\label{eq:p12ExactSimplified}
	p_{12} 
	= \beta \cdot \sum_{k=cn+1}^{dn-1} \sum_{l=k+1}^{dn} \frac{\binom{n-2}{l-2}}{\binom{n-3}{l-3}}
	> \frac{c}{n} \cdot \sum_{k=cn+1}^{dn-1} \sum_{l=k+1}^{dn} \frac{1}{l-2} \,.
	\end{equation}
	Since 
	$ \sum_{l=k+1}^{dn} \frac{1}{l-2} = \sum_{l=k-1}^{dn-2} \frac{1}{l} = \left(\sum_{l=k}^{dn-1} \frac{1}{l}\right) + \frac{1}{k-1} - \frac{1}{dn-1}$
	and $1/l$ is monotonically decreasing, we have $\sum_{l=k}^{dn-1} \frac{1}{l} \geq \int_{k}^{dn} \frac{1}{l} \intD{\ell} = \ln \frac{dn}{k}$ by Fact~\ref{fact:approxSumByIntegralDec}.
	Therefore,
	\begin{align}
	\label{ineq:p12Interim1}
	\sum_{k=cn+1}^{dn-1} \sum_{l=k+1}^{dn} \frac{1}{l-2}
	&\geq \sum_{k=cn+1}^{dn-1} \left( \ln \frac{dn}{k} + \frac{1}{k-1} - \frac{1}{dn-1} \right) \notag \\
	&= \left(\sum_{k=cn+1}^{dn-1} \ln \frac{dn}{k}\right) + \left( \sum_{k=cn+1}^{dn-1} \frac{1}{k-1} \right) - \frac{dn-1-cn}{dn-1} \,.
	\end{align}
	Similarly, using Fact~\ref{fact:approxSumByIntegralDec}, we obtain
	\begin{multline}
	\label{ineq:p12Interim2}
	\sum_{k=cn+1}^{dn-1} \ln \frac{dn}{k}
	= \left(\sum_{k=cn}^{dn-1} \ln \frac{dn}{k}\right) - \ln \frac{d}{c}
	\geq \left(\int_{cn}^{dn} \ln \frac{dn}{k} \intD{k}\right) - \ln \frac{d}{c}
	= dn -cn \cdot \ln \frac{d}{c} -cn - \ln \frac{d}{c}
	\end{multline}
	and 
	\begin{multline}
	\label{ineq:p12Interim3}
	\sum_{k=cn+1}^{dn-1} \frac{1}{k-1}
	= \sum_{k=cn}^{dn-2} \frac{1}{k}
	= \left(\sum_{k=cn}^{dn-1} \frac{1}{k}\right) - \frac{1}{dn-1}
	\geq \left(\int_{cn}^{dn} \frac{1}{k} \intD{k} \right) - \frac{1}{dn-1}
	= \ln \frac{d}{c} - \frac{1}{dn-1} \,.
	\end{multline}
	By combining inequalities (\ref{eq:p12ExactSimplified}) to (\ref{ineq:p12Interim3}), we obtain
	\begin{align*}
	p_{12}
	&> \frac{c}{n} \cdot \left( \left(cn -cn \cdot \ln \frac{d}{c} -cn - \ln \frac{d}{c}\right) + \left( \ln \frac{d}{c} - \frac{1}{dn-1} \right) - \frac{dn-1-cn}{dn-1} \right) \\
	&=c \cdot \left( d - c \ln \frac{d}{c} -c \right) - \frac{c}{n} \cdot \left( 1 - \frac{cn-1}{dn-1} \right) \,.
	\end{align*}
	Since $\frac{c}{n} \cdot \left( 1 - \frac{cn-1}{dn-1} \right) = o(1)$, this gives the claim for $p_{12}$.

	Next, we find a lower bound for $p_{23}$.
	Equation~(\ref{eq:exactPairProbability}) with $j=3$ gives
	\begin{equation}
	p_{23} = 
	\beta \sum_{k=cn+1}^{dn-1} \sum_{l=k+1}^{dn} \frac{\binom{n-3}{l-2}}{\binom{n-3}{l-3}}
	= \beta \cdot \sum_{k=cn+1}^{dn-1} \sum_{l=k+1}^{dn} \frac{n-l}{l-2} \,.
	\end{equation}
	By splitting this expression into two parts we obtain
	\begin{align*}
	p_{23} &= \left(\beta \cdot n \cdot \sum_{k=cn+1}^{dn-1} \sum_{l=k+1}^{dn} \frac{1}{l-2} \right)
					- \beta \cdot \sum_{k=cn+1}^{dn-1} \sum_{l=k+1}^{dn} \frac{l}{l-2} \\
	&> p_{12} - \beta \cdot \sum_{k=cn+1}^{dn-1} \sum_{l=k+1}^{dn} \frac{l}{l-2} \,,
	\end{align*}
	where the inequality follows from inequality~(\ref{eq:p12ExactSimplified}).
	Hence, using the lower bound for $p_{12}$, the claim for $p_{23}$ follows if we can show $\beta \cdot \sum_{k=cn+1}^{dn-1} \sum_{l=k+1}^{dn} \frac{l}{l-2} \leq c \cdot \left( d^2/2 - cd + c^2/2 \right) + o(1)$.
	Since $\frac{l}{l-2}$ decreases monotonically in $l$, Fact~\ref{fact:approxSumByIntegralDec} implies
	\begin{equation}
	\sum_{l=k+1}^{dn} \frac{l}{l-2}
	\leq \int_{k}^{dn} \frac{l}{l-2} \intD{l}
	= dn + 2 \cdot \ln (dn-2) - k - 2 \cdot \ln(k-2) \,.
	\end{equation}
	Therefore, with $\xi = \sum_{k=cn+1}^{dn-1} \ln(k-2)$, we have
	\begin{align*}
	\sum_{k=cn+1}^{dn-1} \sum_{l=k+1}^{dn} \frac{l}{l-2}
	&\leq \sum_{k=cn+1}^{dn-1} \left( dn + 2 \cdot \ln (dn-2) - k - 2 \cdot \ln(k-2) \right) \\
	&= (dn-1-cn) \cdot ( dn + 2 \cdot \ln (dn-2) ) - \left(\sum_{k=cn+1}^{dn-1}  k\right)  - 2 \xi \,.
	\end{align*}
	Since 
	$
	\sum_{k=cn+1}^{dn-1}  k = \frac{(dn-1) \cdot dn}{2} - \frac{cn\cdot (cn+1)}{2}
	$,
	it follows further
	\begin{align*}
	&\sum_{k=cn+1}^{dn-1} \sum_{l=k+1}^{dn} \frac{l}{l-2} \\
	&\leq \frac{(dn-1)\cdot dn}{2} +2(dn-1) \ln(dn-2) \\ 
			&\qquad - cn \cdot \left( dn + 2 \cdot \ln (dn-2)  - \frac{cn+1}{2} \right) - 2\xi \\
	&< \frac{(dn)^2}{2} - n^2cd  + \frac{(cn)^2}{2} + \frac{cn}{2} + 2n(d-c) \cdot \ln(dn-2) - 2\xi \,. \\
	\end{align*}
	Using $\beta = \frac{c}{n-1} \cdot \frac{1}{n-2} < \frac{c}{n^2} \cdot ( 1 + \frac{3}{n-3})$, we get
	\[
	\beta \cdot \sum_{k=cn+1}^{dn-1} \sum_{l=k+1}^{dn} \frac{l}{l-2}
	< c \cdot \left( \frac{d^2}{2} -cd + \frac{c^2}{2} \right) + \eta_1 + \eta_2 \,,
	\]
	where
	$\eta_1 = \frac{3c}{n-3} \cdot \left( \frac{d^2}{2} -cd + \frac{c^2}{2} \right) = o(1)$ and
	\begin{align*}
	\eta_2 
	&= \frac{c}{n^2} \cdot \left( 1 + \frac{3}{n-3} \right) \cdot \left( \frac{cn}{2} + 2n(d-c) \cdot \ln(dn-2) - 2\xi \right) \,.
	\end{align*}	
	We observe that
	\begin{align*}
	\xi &= \sum_{k=cn+1}^{dn-1} \ln(k-2) \\
	&= \left(\sum_{k=cn+1}^{dn} \ln k \right) - \left(\sum_{k=0}^{2} \ln (dn-k) \right) + \ln (cn) + \ln (cn-1) \\
	&\geq  \left(\int_{cn}^{dn} \ln k \intD{k} \right) - \ln \frac{d}{c} - \ln \frac{dn-1}{cn-1} - \ln (dn-2) \\
	&= n \cdot \left( d \ln dn - d - c \ln cn + c \right) - \ln \frac{d}{c} - \ln \frac{dn-1}{cn-1} - \ln (dn-2)
	\end{align*}
	by Fact~\ref{fact:approxSumByIntegralInc}. This implies $\eta_2 = o(1)$ and concludes the proof.
\end{proof}

The remaining probabilities $p_{21}$ and $p_{32}$ can be obtained from the symmetry property stated in the next lemma.

\begin{lemma}
	\label{lemma:2secSymmetry}
	For any two items $i$ and $j$ it holds that	$p_{ij} = p_{ji}$.
\end{lemma}
\begin{proof}
	Suppose $i$ is accepted first and $j$ is accepted as the second item in the input sequence $\pi$.
	Consider the sequence $\pi'$ obtained from $\pi$ by swapping $i$ with $j$.
	Since $j$ and $i$ are the first two elements beating the best sampling item in $\pi'$,
	Algorithm~\ref{alg:2sec} will select $j$ and $i$ on input $\pi'$.
	Hence, the number of permutations must be the same for both events, which implies the claim.
\end{proof}
Therefore, we can obtain all probabilities from Table~\ref{tab:packingTypes} using Lemmas~\ref{lemma:2secSingleItemProbabilitiies}, \ref{lemma:2secPairProbabilities}, and~\ref{lemma:2secSymmetry}.
\subsection{Analysis}
\label{sec:2secCases}
Let $T$ be the set of items in the optimal packing of $I_L$.
This set may contain a single item, may be a two-item subset of $\{1,2,3\}$, or may be a two-item subset containing 
an item $j \geq 4$. In the following, we analyze the performance of Algorithm~\ref{alg:2sec} for each case.

\subsubsection{Single-item case}
If the optimal packing contains a single item, it is the most profitable item. 
Let case 1 be this case. Here, we have $T=\{1\}$ and $\E{\mathcal{A}_L} \geq p_\text{D} \opt_L$.

\subsubsection{Two-item cases}
In cases 2-4, we consider packings of the form $T = \{i,j\}$ with $1 \leq i < j \leq 3$.
We define cases 2, 3, and 4 as $T = \{1,2\}$, $T = \{1,3\}$, and $T = \{2,3\}$, respectively.
We want to consider all algorithmic packings whose profit can be bounded in terms of $\opt_L = v_i + v_j$.
For this purpose, for each case 2-4 we build three groups of feasible packing types, according to whether
the profit of a packing is $\opt_L$, at least $v_i$, or in the interval $[v_j,v_i)$.
We ensure that no packing is counted multiple times by
(a) choosing appropriate packing types and
(b) grouping these packing types in a disjoint way, according to their profit.
Let $\alpha_w$ be the probability that the algorithm returns the optimal packing in case $w \in \{2,3,4\}$.
It holds that $\alpha_2 = p_\text{A}$,  $\alpha_3 = p_\text{B}$, and $\alpha_4 = p_\text{C}$.
In addition, let $\beta_w$ be the probability that an item $k \leq i$ is packed as the first item in case $w \in \{2,3,4\}$.
We have $\beta_2 = p_\text{H}$, $\beta_3 = p_\text{I}$, and $\beta_4 = p_\text{D} + p_\text{K}$.
Finally, let $\gamma_w$ be the probability that an item $k$ with $i < k \leq j$ is packed as the first item in case $w \in \{2,3,4\}$.
It holds that $\gamma_2 = p_\text{J}$, $\gamma_3 = p_\text{E} + p_\text{L}$, and $\gamma_4 = p_\text{M}$.

Finally, we define case 5 as $T = \{i, j\}$ with $i \geq 1$, $j \geq 4$, and $i < j$.
In this case, note that packings of type D contain an item of value at least $v_i$,
and packings of type E, F, and G contain an item of value at least $v_j$.
Hence, we can slightly abuse the notation and set $\alpha_5 = 0$, $\beta_5=p_\text{D}$, and $\gamma_5=p_\text{E}+p_\text{F}+p_\text{G}$,
such that it holds that
\[
\E{\mathcal{A}_L}
\geq \alpha_w (v_i+v_j) + \beta_w v_i + \gamma_w v_j 
\hspace*{20pt}
\text{in case } w \in \{2,3,4,5\} \,.
\]
To bound this term against $\opt_L = v_i + v_j$, consider the following two cases:
If $\beta_w \geq \gamma_w$, we obtain from Chebyshev's sum inequality\footnote{
	Let $a_1 \geq a_2 \geq \ldots \geq a_n$ and $b_1 \geq b_2 \geq \ldots \geq b_n$.
	Chebyshev's sum inequality states that 
	$\sum_{i=1}^{n} a_i b_i \geq (1/n) \left( \sum_{i=1}^{n} a_i\right) \left( \sum_{i=1}^{n} b_i\right)$.}
\[\beta_w v_i + \gamma_w v_j \geq \frac{1}{2} \left( \beta_w + \gamma_w \right) (v_i + v_j) \,. \]
If $\beta_w < \gamma_w$, we trivially have $\beta_w v_i + \gamma_w v_j > \beta_w (v_i + v_j)$.

\subsubsection{Competitive ratio}
The competitive ratio of $\mathcal{A}_L$ is the minimum over all cases 1-5.
Hence, setting $\alpha_1=p_\text{D}$ and $\beta_1 = \gamma_1 = 0$, we obtain
\begin{equation}
\label{eq:2secCaseCompetitiveRatio}
\E{\mathcal{A}_L} 
\geq \min_{w=1,\ldots,5} \left \{
\alpha_w + \min \left\{  \frac{\beta_w + \gamma_w }{2} , \beta_w \right\} \right\} \cdot \opt_L \,.
\end{equation}
Clearly, inequality~(\ref{eq:2secCaseCompetitiveRatio}) simplifies depending on $\beta_w \geq \gamma_w$ or $\beta_w < \gamma_w$. The following lemma gives a sufficient condition for $\beta_w \geq \gamma_w$.

\begin{lemma}
	\label{lemma:2SecFulfillsTechnicalAssumptions}
	Let $f(x)=2 \ln x -6x + 2x^2 - \frac{x^3}{3}$.
	For parameters $c$, $d$ with $f(c) \geq f(d)$ and $n \to \infty$, it holds that $\beta_w \geq \gamma_w$,
	where $2 \leq w \leq 5$.
\end{lemma}
\begin{proof}
	We first show that $f(c) \geq f(d)$ is equivalent to $\beta_5 \geq \gamma_5$.
	Note that $\beta_5 = p_\text{D} = p_1$ and $\gamma_5 = p_\text{E} + p_\text{F} + p_\text{G} = p_2 + p_3 + p_4$.
	Now, using Lemma~\ref{lemma:2secSingleItemProbabilitiies} and ignoring lower order terms, we have
	\begin{align*}
		&& p_1 & \geq p_2 + p_3 + p_4 \\
		\Leftrightarrow && c \ln \frac{d}{c} & \geq c \left(3 \ln \frac{d}{c} - 6(d-c) + 2(d^2-c^2) - \frac{1}{3} (d^3-c^3) \right) \\
		\Leftrightarrow  && 0 & \geq 2 \ln \frac{d}{c} - 6(d-c) + 2(d^2-c^2) - \frac{1}{3} (d^3-c^3) \\		
		\Leftrightarrow  && 0 & \geq 2 \ln d - 2 \ln c  - 6d + 6c + 2 d^2 -2c^2 - \frac{d^3}{3} + \frac{c^3}{3} \\		
		\Leftrightarrow  && f(c) & \geq f(d) \,.
	\end{align*}
	Therefore, the claim for $w=5$ holds by assumption.
	For $2 \leq w \leq 4$, 
	the claims follow immediately from $f(c) \geq f(d)$
	and the symmetry property of Lemma~\ref{lemma:2secSymmetry}:
	\begin{align*}
		\beta_2 =&~ p_\text{H} = p_1 - p_{12} = p_1 - p_{21} \geq p_2 - p_{21} = p_\text{J} = \gamma_2 \\
		\beta_3 =&~ p_\text{I} = p_1 - p_{13} = p_1 - p_{31} \geq p_2 + p_3 - p_{31} = p_E + p_\text{L} = \gamma_3 \\
		\beta_4 =&~ p_\text{D} + p_\text{K} = p_1 + p_2 - p_{23} \geq p_1 - p_{32} \geq p_3 - p_{32} = p_\text{M} = \gamma_4 \,.
	\end{align*}
\end{proof}

We obtain the following two lemmas.
If $\mathcal{A}_L$ uses the entire input sequence ($d=1$), this algorithm is ($1 / 3.08$)-competitive.
\begin{lemma}
	\label{lemma:2secD1}
	With $c = 0.23053$ and $d = 1$ as parameters, we have $\E{\mathcal{A}_L} \geq \left( \frac{1}{3.08} -o(1) \right) \cdot \opt_L$.
\end{lemma}

\noindent
Note that 2-KS includes the secretary problem (case 1); thus, no algorithm for 2-KS can have a better competitive ratio than $1/e < 1/2.71$.
In the final algorithm we set $d<1$ to benefit from $\mathcal{A}_S$.
The next lemma has already been used to prove Theorem~\ref{thm:KSmainTheorem} in Section~\ref{sec:sequentialApproach}.
\begin{lemma}
	\label{lemma:2secFinalParameters}
	With $c = 0.42291$ and $d = 0.64570$ as parameters, we have $\E{\mathcal{A}_L} \geq \left( \frac{1}{6.65} - o(1) \right) \cdot \opt_L$.
\end{lemma}

\begin{table}
	\begin{center}
		\caption{Competitive ratios of Algorithm~\ref{alg:2sec} for the parameters from Lemmas \ref{lemma:2secD1} and \ref{lemma:2secFinalParameters} in different cases. 
			Bold values indicate the minimum over all cases and thus the competitive ratio.}
		\label{tab:2SecPerformance}		
		\begin{tabular}{llllllll} \toprule
			&	& & & \multicolumn{4}{c}{Two-item cases}\\ \cmidrule{5-8}
			&	$c$ & $d$ & Case 1 & Case 2 & Case 3 & Case 4 & Case 5 \\ \midrule
			
			Lemma~\ref{lemma:2secD1} & 0.23053 & 1  & 0.33827  & 0.34898 & 0.32705 & 0.32705 & \textbf{0.32471}  \\ 
			Lemma~\ref{lemma:2secFinalParameters} & 0.42291 & 0.64570 & 0.17897  & \textbf{0.15039} & 0.16033 & 0.16033 & 0.16231   \\ \midrule
		\end{tabular} 
	\end{center}
\end{table}

\begin{proof}[Proof of Lemmas \ref{lemma:2secD1} and \ref{lemma:2secFinalParameters}]
	Let $f$ be the function defined in Lemma~\ref{lemma:2SecFulfillsTechnicalAssumptions}
	and let $(c_1,d_1) = (0.23053,1)$ and $(c_2,d_2) = (0.42291,0.64570)$ be the two parameter pairs from Lemmas~\ref{lemma:2secD1} and~\ref{lemma:2secFinalParameters}, respectively. It holds that
	\[
	f(c_1)  = f(0.23053) > -4.22 > -\frac{13}{3} = f(1) = f(d_1)
	\]
	and
	\[
	f(c_2)  = f(0.42291) > -3.93 > -4.00 > f(0.64570) = f(d_2) \,.
	\]
	Hence, by Lemma~\ref{lemma:2SecFulfillsTechnicalAssumptions} we have $\beta_w \geq \gamma_w$ for any case $w \in \{2,3,4,5\}$.
	Therefore, inequality~(\ref{eq:2secCaseCompetitiveRatio}) simplifies to
	$\E{\mathcal{A}_L} 	\geq \min_{w=1,\ldots,5}  \left \{ \alpha_w + \frac{\beta_w + \gamma_w }{2}\right \} \cdot \opt_L$.
	Using the definitions of $\alpha_w$, $\beta_w$, and $\gamma_w$ from Section~\ref{sec:2secCases}, the definitions of
	$p_X$ from Table~\ref{tab:packingTypes}, and the symmetry property of Lemma~\ref{lemma:2secSymmetry}, we obtain after simplifying terms
	\begin{align*}	
	\alpha_2 + \frac{\beta_2 + \gamma_2}{2} 
	&= p_\text{A} + \frac{p_\text{H} + p_\text{J}}{2}
	= \frac{p_1 + p_2}{2} + p_{12} \\
	\alpha_3 + \frac{\beta_3 + \gamma_3}{2} 
	&= p_\text{B} + \frac{p_\text{I} + (p_\text{E} + p_\text{L})}{2}
	= \frac{p_1 + p_2 + p_3}{2} + p_{13} \\
	\alpha_4 + \frac{\beta_4 + \gamma_4}{2} 
	&= p_\text{C} + \frac{(p_\text{D} + p_\text{K}) + p_\text{M}}{2}
	= \frac{p_1 + p_2 + p_3}{2} + p_{23} \\
	\alpha_5 + \frac{\beta_5 + \gamma_5}{2} 
	&= 0 + \frac{p_\text{D} + (p_\text{E} + p_\text{F} + p_\text{G})}{2}
	= \frac{p_1 + p_2 + p_3 + p_4}{2} \,.
	\end{align*}
	Note that the algorithm attains the same competitive ratio in case 3 and 4, since $p_{13} = p_{23}$ by Lemma~\ref{lemma:2secPairProbabilities}.
	Table~\ref{tab:2SecPerformance} shows the competitive ratios for all five cases.
	For the overall competitive ratio, we have
	\[
	\E{\mathcal{A}_L} \geq \min \left\{ p_1, p_{12} + \frac{p_1 + p_2}{2}, p_{23} + \frac{p_1 + p_2 + p_3}{2}, \frac{\sum_{i=1}^4 p_i}{2} \right\} \opt_L \,.
	\]
	Evaluating this expression for the parameter pairs $(c_1,d_1)$ and $(c_2,d_2)$ yields
	$0.32471 \geq 1/3.08$ and $0.15039 \geq 1/6.65$ as competitive ratios, respectively. This concludes the proofs of Lemmas~\ref{lemma:2secD1} and~\ref{lemma:2secFinalParameters}.
\end{proof}

Recall that in Algorithm~\ref{alg:sequentialAlg}, we can only benefit from $\mathcal{A}_S$ if $\mathcal{A}_L$
has not filled the knapsack completely.
Thus, the following property is crucial in the final analysis.
\begin{lemma}
	\label{lemma:2SecEmptyKnapsack}
	With a probability of at least $c/d$, no item is packed by $\mathcal{A}_L$.
\end{lemma}
\begin{proof}
	Fix any set of $dn$ items arriving in rounds $1,\ldots,dn$.
	The most profitable item $v^*$ from this set arrives in the sampling phase with probability $c/d$.
	If this event occurs, no item in rounds $cn+1,\ldots,dn$ beats $v^*$ and
	$\mathcal{A}_L$ will not select any item.
\end{proof}

\subsection{Discussion of other 2-Secretary Algorithms}
\label{sec:kSecAlgorithms}

As mentioned in the introduction of Section~\ref{sec:largeItems}, the approach and its analysis of this section are general enough to cover all two-choice secretary algorithms. Therefore, a natural question to ask is which algorithm is a good choice within this framework. Algorithm~\ref{alg:2sec} is based on the algorithm \textsc{single-ref} developed for the $k$-secretary problem in \cite{kSecretary}. In the following, we discuss several algorithms for $2$-secretary and related problems.

The \textsc{optimistic} algorithm by Babaioff et al.\ \cite{DBLP:conf/approx/BabaioffIKK07} was developed for the $k$-secretary problem and performs slightly better than \textsc{single-ref} in the case $k=2$;
the competitive ratios of both algorithms are $0.4168$ and $0.4119$, respectively \cite{kSecretary}.
However, \textsc{optimistic} has a weaker threshold for accepting the first item than \textsc{single-ref},
thus the probability considered in Lemma~\ref{lemma:2SecEmptyKnapsack} would fall below $c/d$.
In the present analysis of the sequential approach, we can only benefit from the second algorithm $\mathcal{A}_S$ if $\mathcal{A}_S$ starts with an empty knapsack (we will use this property later in Lemma~\ref{lemma:KSsmallItemsFinal}).
Hence, it is not clear if the slight gain in the expected profit compensates the drawback of an early resource consumption.

A strong algorithm for the 2-secretary problem has been developed by Chan et al.\ \cite{DBLP:conf/soda/ChanCJ15}. The algorithm is based on a sophisticated set of decision rules, leading to a competitive ratio of $0.49$.
Again, the probability considered in Lemma~\ref{lemma:2SecEmptyKnapsack} would be smaller for this algorithm. Moreover, it seems overly elaborate to find equivalents of Lemmas~\ref{lemma:2secSingleItemProbabilitiies}, \ref{lemma:2secPairProbabilities}, \ref{lemma:2SecFulfillsTechnicalAssumptions}, and \ref{lemma:2SecEmptyKnapsack}.

Another candidate algorithm is due to Nikolaev \cite{nikolaev1977generalization} and 
Tamaki \cite{tamaki1979recognizing} who proposed an algorithm for a slightly different secretary problem:
Here, the objective is to maximize the probability of selecting the best two items.
This algorithm depends on two parameters $0 \leq c_1 \leq c_2 \leq 1$.
The first item is selected just as in \textsc{single-ref} with sampling size $c_1 n$ (select the first item beating
the best sampling item). The second item must beat the first item if it arrives before round $c_2 n$,
or (merely) the best sampling item if it arrives later than this round.
The success probability tends asymptotically to $0.2254$ with $c_1 = 0.2291$ and $c_2=0.6065$, which is best possible \cite{tamaki1979recognizing}.
If we use this algorithm within our framework, it turns out that the best competitive ratio is achieved for $c_1 = c_2$.
However, for $c_1 = c_2$, this algorithm is equal to \textsc{single-ref} in the case $k=2$.

Therefore, we conclude that even though various algorithms for the 2-secretary problem stronger than \textsc{single-ref} exist, it is not clear if they can improve the performance of the overall algorithm within the sequential framework. On the other side, Algorithm~\ref{alg:2sec} (based on \textsc{single-ref}) is fairly easy to analyze and selects high-profit items with sufficient high probability.

\section{Small Items}
\label{sec:smallItems}

For $(1/3)$-small items, we use solutions for the fractional problem variant and obtain an integral packing
via randomized rounding. This approach has been applied successfully to packing LPs \cite{DBLP:journals/siamcomp/KesselheimRTV18};
however, for the knapsack problem it is not required to solve LP relaxations in each round (as in \cite{DBLP:journals/siamcomp/KesselheimRTV18}).
Instead, here, we use solutions of a greedy algorithm, which is well-known to be optimal for the fractional knapsack problem. Particularly, this algorithm is both efficient in running time and easy to analyze.

We next formalize the greedy solution for any set $T$ of items.
Let the \textit{density} of an item be the ratio of its profit to its size.
Consider any list $L$ containing the items from $T$ ordered by non-increasing density.
We define the \textit{rank} $\rho (i)$ of item $i$ as its position in $L$
and $\sigma(l)$ as the item at position $l$ in $L$. Thus, $\sigma(l) = \rho^{-1}(l)$ denotes the $l$-th densest item.
Let $k$ be such that $\sum_{i=1}^{k-1} s_{\sigma(i)} < W \leq \sum_{i=1}^{k} s_{\sigma(i)}$.
The fraction of item $i$ in the greedy solution $\alpha$ is now defined as
\[
\alpha_i = 
\begin{cases}
1 & \text{if } \rho(i) < k \\
\left(W - \sum_{i=1}^{k-1} s_{\sigma(i)}\right) / s_i & \text{if } \rho(i) = k \\
0 & \text{else} \,,
\end{cases}
\]
i.e., the $k-1$ densest items are packed integrally and the remaining space is filled by the maximum feasible fraction of the $k$-th densest item.
Let $\opt(T)$ and $\opt^*(T)$ denote the profits of optimal integral and fractional packings of $T$, 
respectively. It is easy to see that $\alpha$ satisfies
$\sum_{i \in T} \alpha_i v_i = \opt^*(T) \geq \opt(T) $ and $\sum_{i \in T} \alpha_i s_i = W$.

\subsection{Algorithm}
The algorithm $\mathcal{A}_S$ for $(1/3)$-small items, which is formally defined in Algorithm~\ref{alg:smallItems},
works as follows. During the initial sampling phase of $dn$ rounds, the algorithm rejects all items.
In each round $\ell \geq dn+1$, the algorithm 
computes a greedy solution $x^{(\ell)}$ for $I_S(\ell)$. Here, 
$I_S(\ell)$ denotes the subset of $I_S$ revealed up to round $\ell$.
The algorithm packs the current online item $i$ with probability $x^{(\ell)}_i$.
However, generally, this can only be done if the remaining capacity of the knapsack is at least $(1/3) \cdot W \geq s_i$.

Note that in case of an integral coefficient $x^{(\ell)}_i \in \{0,1\}$, the packing step is completely deterministic. Moreover,
in any greedy solution $x^{(\ell)}$, there is at most one item $i$ with fractional coefficient $x^{(\ell)}_i \in (0,1)$. Therefore, in expectation, there is only a small number of rounds where the algorithm actually requests randomness. Although this is not relevant for the proof of the competitive ratio, we provide a short proof of this observation in the following.

\begin{observation}
	\label{lemma:lpAlgRandomness}
	Let $X$ denote the number of rounds where Algorithm~\ref{alg:smallItems} packs an item with probability $x_i \in (0,1)$. It holds that $\E{X} \leq \ln (1/d) \leq 0.44$.
\end{observation}
\begin{proof}
	Consider any round $\ell$ and let $x^{(\ell)}$ be the greedy knapsack solution computed by Algorithm~\ref{alg:smallItems}. By definition of $x^{(\ell)}$, at most one of the $\ell$ visible items has a fractional coefficient $x^{(\ell)}_i \in (0,1)$. The probability that this item $i$ arrives in round $\ell$ is $1/\ell$ in a random permutation.
	Let $X_\ell$ be an indicator variable for the event that Algorithm~\ref{alg:smallItems} packs an item at random in round $\ell$. By the above argument, we have $\Pr{X_\ell = 1} \leq 1/\ell$. Since Algorithm~\ref{alg:smallItems} selects items starting in round $dn+1$, we obtain
	\[
	\E{X} = \sum_{\ell=dn+1}^{n} \E{X_\ell} \leq \sum_{\ell=dn+1}^{n} \frac{1}{\ell} \leq \ln \frac{1}{d} \leq 0.44 \,.
	\qedhere
	\]
\end{proof}
Note that Algorithm~\ref{alg:2sec} and the sequential approach (Algorithm~\ref{alg:sequentialAlg}) are deterministic algorithms. Therefore, our overall algorithm requests randomness in expectation in less than one round.

\begin{algorithm}[t]
	\SetKwInOut{Input}{Input}
	\SetKwInOut{Output}{Output}
	\Input{Random permutation of $n$ $(1/3)$-small items, a knapsack of capacity $W$,\\
		parameter $d \in (0,1)$.}
	\Output{A feasible (integral) packing of the knapsack.}
	Let $\ell$ be the current round and $i$ be the online item of round $\ell$. \\
	\uIf{$\ell \leq dn$} {Sampling phase -- reject all items.}
	\uIf{$dn+1 \leq \ell \leq n$} {
		Let $x^{(\ell)}$ be the greedy solution for $I_S (\ell)$. \\
		\uIf{the remaining capacity is at least $(1/3)\cdot W$} {
			Pack $i$ with probability $x^{(\ell)}_i$.
		}
 	}	
	\caption{Algorithm $\mathcal{A}_S$ for small items}
	\label{alg:smallItems}
\end{algorithm}

\subsection{Analysis}
\label{sec:ksSmallAnalysis}
Before we analyze the competitive ratio of $\mathcal{A}_S$ in a sequence of lemmas, we make a few technical observations and introduce further notation.

In round $dn+1$, the knapsack might already have been filled by $\mathcal{A}_L$ with large items 
from previous rounds. For now, we assume an empty knapsack after round $dn$ and denote this event by $\xi$.
In the final analysis, we will use the fact that $\Pr{\xi}$ can be bounded from below, which is according to  
Lemma~\ref{lemma:2SecEmptyKnapsack}.

The description of Algorithm~\ref{alg:smallItems} is tailored to $(1/3)$-small items, in order to complement Algorithm~\ref{alg:2sec}. Anyway, it is straightforward to generalize this algorithm to arbitrary maximum item size $\delta$. In order to show similarities with the analysis from Section~\ref{sec:gap} later, we state the following lemmas with $\delta$ as a parameter. For this purpose, we define $\Delta = \frac{1}{1-\delta}$ (and obtain $\Delta=3/2$ in the final analysis).

Finally, let $\alpha$ be a greedy (offline) solution for $I_S$. By the following lemma, the probability that an item $i \in I_S$ is packed by $\mathcal{A}_S$ is proportional to $\alpha_i$. By treating $\alpha_i$ as a parameter in the next two lemmas, it is not required to analyze the profit in each round in expectation over all items. The latter approach appears in related work \cite{DBLP:conf/esa/KesselheimRTV13}, where stochastic dependencies need to be handled carefully.

\begin{lemma}
	\label{lemma:knapsackItemPackedSingleRound}
	Let $i \in I_S$ and $E_i(\ell)$ be the event that the item $i$ is packed by $\mathcal{A}_S$ in round $\ell$.
	For $\ell \geq dn+1$, it holds that
	$\Pr{E_i(\ell) \mid \xi} \geq \frac{1}{n} \alpha_i (1 - \Delta \ln \frac{\ell}{dn})$.
\end{lemma}
\begin{proof}
	In a random permutation, item $i$ arrives in round $\ell$ with probability $1/n$.
	In round $\ell \geq dn+1$, the algorithm decides to pack $i$ with probability $x^{(\ell)}_i$.
	Note that the rank of item $i$ in $I_S(\ell)$ is less than or equal to its rank in $I_S$.
	According to the greedy solution's definition, this implies $x^{(\ell)}_i \geq \alpha_i$.
	Finally, the $\delta$-small item $i$ can be packed successfully
	if the current resource consumption $X$ is at most $(1 - \delta) W$.
	In the following, we investigate the expectation of $X$ to give a probability bound using Markov's inequality at the end of this proof.
	
	Let $X_k$ be the resource consumption in round $k < \ell$.
	By assumption, the knapsack is empty after round $dn$, thus $X = \sum_{k=dn+1}^{\ell-1} X_k$.
	Let $Q$ be the set of $k$ visible items in round $k$.
	The set $Q$ can be seen as uniformly drawn from all $k$-item subsets and any item $j \in Q$ is the current
	online item of round $k$ with probability $1/k$.
	The algorithm packs any item $j$ with probability $x_j^{(k)}$, thus
	\begin{equation*}
		\E{X_k}
		= \sum_{j \in Q} \Pr{j \text{ occurs in round } k} s_j x_j^{(k)}
		= \frac{1}{k} \sum_{j \in Q} s_j x_j^{(k)}
		\leq \frac{W}{k} \,,
	\end{equation*}
	where the last inequality holds because $x^{(k)}$ is a feasible solution for a knapsack of size $W$.
	By the linearity of expectation and the previous inequality,
	the expected resource consumption up to round $\ell$ is
	\[\E{X} = \sum_{k=dn+1}^{\ell-1} \E{X_k} \leq \sum_{k=dn+1}^{\ell-1} \frac{W}{k} \leq W \ln \frac{\ell}{dn} \,.
	\]
	Using Markov's inequality, we obtain
	\begin{equation*}
		\Pr{X < (1-\delta) W}  
		= 1 - \Pr{X \geq (1-\delta) W}
		\geq 1 - \frac{\E{X}}{(1-\delta)W}
		\geq 1 - \Delta \ln \frac{\ell}{dn} \,,
	\end{equation*}
	which concludes the proof.
\end{proof}
Using Lemma~\ref{lemma:knapsackItemPackedSingleRound} we easily obtain the total probability that
a specific item will be packed.
\begin{lemma}
	\label{lemma:knapsackItemPacked}
	Let $i \in I_S$ and $E_i$ be the event that the item $i$ is packed by $\mathcal{A}_S$. It holds that
	$\Pr{E_i \mid \xi} \geq \alpha_i \left( (1-d)(1+\Delta) - \Delta \cdot \left( 1 + \frac{1}{n} \right) \cdot \ln \frac{1}{d} \right) $.
\end{lemma}
\begin{proof}
	Summing the probabilities from Lemma~\ref{lemma:knapsackItemPackedSingleRound} over all rounds $\ell \geq dn+1$ gives
	\begin{align}
		\Pr{E_i \mid \xi}
		&= \sum_{\ell=dn+1}^{n} \Pr{E_i(\ell) \mid \xi}   \notag \\
		&\geq \sum_{\ell=dn+1}^{n} \frac{1}{n} \alpha_i \left( 1 - \Delta \ln \frac{\ell}{dn} \right) \notag \\
		&= \frac{1}{n} \alpha_i \left( n - dn - \Delta \sum_{\ell=dn+1}^{n}  \ln \frac{\ell}{dn} \right) \notag \\
		&= \alpha_i \left( 1 - d - \frac{\Delta}{n} \sum_{\ell=dn+1}^{n}  \ln \frac{\ell}{dn} \right) \label{ineq:KSpackingProbApproach} \,.
	\end{align}
	By Fact~\ref{fact:approxSumByIntegralInc}, we obtain
	\begin{align*}
	\sum_{\ell=dn+1}^{n} \ln \frac{\ell}{dn}
	= \left(\sum_{\ell=dn}^{n-1} \ln \frac{\ell}{dn}\right) + \ln \frac{1}{d}
	\leq \left(\int_{dn}^{n} \ln \frac{\ell}{dn} \, \mathrm{d}\ell\right) + \ln \frac{1}{d}
	\end{align*}
	and resolving the integral yields
	\begin{align}
	\sum_{\ell=dn+1}^{n} \ln \frac{\ell}{dn}
	&\leq n \cdot \left( \ln \frac{n}{dn}  - 1 \right) - dn \cdot \left( \ln \frac{dn}{dn} - 1 \right) + \ln \frac{1}{d} \notag \\
	&= n \cdot \ln \frac{1}{d} - n + dn + \ln \frac{1}{d}  \label{ineq:KSpackingProbSum} \,.
	\end{align}
	The claim follows by combining inequalities (\ref{ineq:KSpackingProbApproach}) and (\ref{ineq:KSpackingProbSum}) and by rearranging terms.
\end{proof}
The following lemma bounds the expected profit of the packing of $\mathcal{A}_S$,
assuming the event $\xi$.
\begin{lemma}
	\label{lemma:knapsackTotalProfit}
	We have 
	$\E{\mathcal{A}_S \mid \xi} \geq \left( (1 \!- \! d) (1 \!+ \! \Delta) - \Delta \cdot \left( 1 \!+ \! \frac{1}{n} \right) \cdot \ln \frac{1}{d} \right) \opt_S$.
\end{lemma}
\begin{proof}
	Let $\beta = (1 - d) (1 + \Delta) - \Delta \cdot \left( 1 + \frac{1}{n} \right) \cdot \ln \frac{1}{d}$.
	By Lemma~\ref{lemma:knapsackItemPacked}, the probability that an item $i$ is packed, assuming $\xi$, is
	$\Pr{E_i \mid \xi} \geq \alpha_i \beta$. Therefore,
	\[
	\E{\mathcal{A}_S \mid \xi }
	= \sum_{i \in I_S} \Pr{E_i \mid \xi} v_i
	\geq \sum_{i \in I_S} \alpha_i \beta v_i
	\geq \beta \opt_S \,.
	\qedhere
	\]

\end{proof}
The conditioning on $\xi$ can be resolved using Lemma~\ref{lemma:2SecEmptyKnapsack}.
We obtain the following lemma, which is the second pillar in the proof of Theorem~\ref{thm:KSmainTheorem} and concludes this section.

\begin{lemma}
	\label{lemma:KSsmallItemsFinal}
	We have
	$
	\E{\mathcal{A}_S} 
	\geq \frac{c}{d} \left( (1 - d) (1 + \Delta) - \Delta \cdot \left( 1 + \frac{1}{n} \right) \cdot \ln \frac{1}{d} \right) \opt_S$.
	In particular, the algorithm $\mathcal{A}_S$ is $(1/6.65)$-competitive with respect to $\opt_S$
	setting $\Delta = 3/2$, $c = 0.42291$, and $d = 0.64570$.
	
\end{lemma}
\begin{proof}
	By Lemma~\ref{lemma:2SecEmptyKnapsack}, the probability for an empty knapsack after round $dn$
	is $\Pr{\xi} \geq \frac{c}{d}$. Thus, we obtain from Lemma~\ref{lemma:knapsackTotalProfit} 
	\begin{align*}
	\E{\mathcal{A}_S} 
	&= \Pr{\xi} \E{\mathcal{A}_S \mid \xi} \\
	&\geq \frac{c}{d} \left( (1 - d) (1 + \Delta) - \Delta \cdot \left( 1 + \frac{1}{n} \right) \cdot \ln \frac{1}{d} \right) \opt_S \,.
	\end{align*}
	Setting $\Delta = 3/2$, which corresponds to $\delta=1/3$, leads to 
	\[
	\E{\mathcal{A}_S} 
	\geq \frac{c}{d} \left( \frac{5}{2} (1 - d) - \frac{3}{2} \ln \frac{1}{d} - \frac{3}{2n} \ln \frac{1}{d}\right) \opt_S \,.
	\]
	Noting that $\frac{c}{d} \frac{3}{2n} \ln \frac{1}{d} = o(1)$, we obtain that $\E{\mathcal{A}_S} \geq \left( \frac{1}{6.65} - o(1) \right) \opt_S$
	for $c = 0.42291$ and $d = 0.64570$.
\end{proof}

\section{Extension to GAP}
\label{sec:gap}

In this section, we show that the sequential approach introduced in Section~\ref{sec:sequentialApproach} can be easily adapted to GAP, yielding a $(1/6.99)$-competitive randomized algorithm. We first define the problem formally.

\paragraph{GAP.}
We are given a set of items $I = [n]$ and a set of resources $R = [m]$ 
of capacities $W_r \in \QQ_{> 0}$ for $r \in R$.
If item $i \in I$ is assigned to resource $r \in R$, this raises profit (value) $v_{i,r} \in \QQ_{\geq 0}$, 
but consumes $s_{i,r} \in \QQ_{> 0}$ of the resource's capacity.
The goal is to assign each item to at most one resource such that the total profit is maximized and no resource
exceeds its capacity.
We call the tuple $(v_{i,r}, s_{i,r})$ an \textit{option} of item $i$ and w.l.o.g.\ assume that options for all resources exist. This can be ensured by introducing dummy options with $v_{i,r} = 0$.
In the online version of the problem, in each round an item is revealed together with its set of options. The online algorithm must decide immediately and irrevocably, if the item is assigned. If so, it has to specify the resource according to one of its options.

Again, we construct restricted instances $\mathcal{I}_L$ and $\mathcal{I}_S$ according to the following definition,
which generalizes Definition~\ref{def:itemSizes}. Let $\delta \in (0,1)$.
\begin{definition}
	\label{def:GAPitemSizes}
	We call an option $(v_{i,r}, s_{i,r})$ \textit{$\delta$-large} if $s_{i,r} > \delta W_r$ and \textit{$\delta$-small} 
	if $s_{i,r} \leq \delta W_r$.
	Whenever $\delta$ is clear from the context, we say an option is \textit{large} or \textit{small} for short.
	Based on a given instance $\mathcal{I}$ for GAP, we define two modified instances $\mathcal{I}_L$ and $\mathcal{I}_S$ which are
	obtained from $\mathcal{I}$ as follows.
	\begin{itemize}
		\item $\mathcal{I}_L$: Replace each small option $(v_{i,r}, s_{i,r})$ by the large option $(0,W_r)$.
		\item $\mathcal{I}_S$: Replace each large option $(v_{i,r}, s_{i,r})$ by the small option $(0, \delta W_r)$.	
	\end{itemize}
\end{definition}
Thus, $\mathcal{I}_L$ only contains large options and $\mathcal{I}_S$ only contains small options.
However, by construction no algorithm will assign an item according to a zero-profit option.
We define $\opt$, $\opt_L$, and $\opt_S$ accordingly.
Note that the inequality $\opt \leq \opt_L + \opt_S$ holds also for GAP.

The sequential framework of Algorithm~\ref{alg:sequentialAlg} can be adapted in a straightforward manner by replacing terms like \textit{packing} with \textit{assignment to resource $r$}.
Here, we set the threshold parameter to $\delta=1/2$. In the following subsections, we specify algorithms $\mathcal{A}_L$ and $\mathcal{A}_S$ for $(1/2)$-large and $(1/2)$-small options, respectively.

\subsection{Large Options}
\label{sec:gapLarge}
If each item consumes more than one half of a resource, no two items can be assigned to this resource. 
Thus, we obtain the following matching problem.

\paragraph{Edge-weighted bipartite matching.}
Given a bipartite graph $G=(L \cup R, E)$ and a weighting function $w \colon E \to \QQ_{\geq 0}$,
the goal is to find a bipartite matching $M \subseteq E$ such that $w(M) := \sum_{e \in M} w(e)$ is maximal.
In the online version, the (offline) nodes from $R$ and the number $n=|L|$ are known in advance, 
whereas the nodes from $L$ are revealed online together with their incident edges.
In the case of GAP, $L$ is the set of items, $R$ is the set of resources, and the weight of an edge $e=\{l,r\}$ is $w(e)=v_{l,r}$.

Kesselheim et al.\ \cite{DBLP:conf/esa/KesselheimRTV13} developed an optimal $(1/e)$-competitive algorithm for the online problem under random arrival order.
Adapting this algorithm to the sequential approach with parameters $c$ and $d$ leads to the following algorithm $\mathcal{A}_L$:
During the first $cn$ rounds, no edge is added to the matching.
Then, in each round $\ell$, the algorithm computes a maximum 
edge-weighted matching $M^{(\ell)}$ for the graph revealed up to this round. 
Let $l \in L$ be the online vertex of round $\ell$.
If $l$ is matched in $M^{(\ell)}$ to some node $r \in R$,
we call $e^{(\ell)} = \{l,r\}$ the \textit{tentative edge} of round $\ell$.
Now, if $r$ is still unmatched and $\ell \leq dn$, the tentative edge is added to the matching.

\begin{algorithm}
	\SetKwInOut{Input}{Input}
	\SetKwInOut{Output}{Output}
	\Input{Offline vertex set $R$, number of online vertices $n = |L|$, \\
		parameters $c,d \in (0,1)$ with $c<d$.}
	\Output{Matching $M$.}
	Set $M=\emptyset$. \\
	Let $\ell$ be the current round and $l$ be the online vertex of round $\ell$. \\
	\uIf{$1 \leq \ell \leq cn$} {Sampling phase -- do not add any edge.}
	\uIf{$cn+1 \leq \ell \leq dn$} {
		Let $M^{(\ell)}$ be a maximum-weight matching for the graph in round $\ell$. \\
		Let $e^{(\ell)} \in M^{(\ell)}$ be the edge incident to $l$. \\
		\uIf{$M \cup e^{(\ell)}$ is a matching} {
			Add $e^{(\ell)}$ to $M$.
		}
		\uIf{$\ell > dn$} {Do not add any edge.}
	}
	\caption{Algorithm for edge-weighted bipartite matching from \cite{DBLP:conf/esa/KesselheimRTV13}
		(extended by parameters $c$, $d$).}
	\label{alg:matching}
\end{algorithm}

A formal description of this algorithm is given in Algorithm~\ref{alg:matching}.
The proof of the approximation guarantee relies mainly on the following two lemmas; for completeness, we give the proofs from \cite{DBLP:conf/esa/KesselheimRTV13} here.
The first lemma shows that the expected weight of any tentative edge can be bounded from below.

\begin{lemma}[\cite{DBLP:conf/esa/KesselheimRTV13}]
	\label{lemma:MatchingExpectedWeight}	
	In any round $\ell$, the tentative edge (if it exists) has expected weight $\E{w(e^{(\ell)})} \geq \frac{1}{n} \opt_L$.
\end{lemma}
\begin{proof}
	
	We use the fact that the random sequence of visible items in round $\ell$ can be obtained from the following process: First, the set $Q$ of visible items in round $\ell$ is drawn uniformly at random from all $\ell$-element subsets of $L$. Then, the online vertex of round $\ell$ is drawn uniformly at random from $Q$. Note that these random experiments are independent.

	After the first step, the matching $M^{(\ell)}$ is already fixed.
	Let $M^* = M^{(n)}$ be a maximum weight (offline) matching and 
	$M^*_Q = \{e = \{l,r\} \in M^* \mid l \in Q\}$ the matching $M^*$ projected to visible nodes.
	We have	$w(M^{(\ell)}) \geq w(M^*_Q)$, since $M^{(\ell)}$ is an optimal and $M^*_Q$ a feasible matching for the graph revealed in round $\ell$.
	As described above, each vertex $l \in L$ has probability $\ell/n$ to be in $Q$, thus
	\begin{equation}
	\label{eq:matchingWeightRoundL}
	\E{w(M^{(\ell)})} \geq \E{w(M^*_Q)} = \sum_{e=\{l,r\} \in M^*} \Pr{l \in Q} w(e) = \frac{\ell}{n} w(M^*) \,.
	\end{equation}
	
	For the second step, we observe that each vertex from $Q$ has the same probability of $1/\ell$ to arrive in round $\ell$. 
	Let $\mathcal{M}$ be the domain of the random variable $M^{(\ell)}$.
	We have
	\begin{align}	
	\label{eq:matchingWeightTentativeEdge}
	\E{w(e^{(\ell)})} 
	&= \sum_{M' \in \mathcal{M}} \E{w(e^{(\ell)}) \mid M^{(\ell)} = M'} \cdot \Pr{M^{(\ell)} = M'} \notag \\
	&= \sum_{M' \in \mathcal{M}} \left( \sum_{e=\{l,r\} \in M'} \frac{1}{\ell} w(e) \right) \cdot \Pr{M^{(\ell)} = M'} \notag \\
	&= \frac{1}{\ell} \cdot \sum_{M' \in \mathcal{M}} w(M') \cdot \Pr{M^{(\ell)} = M'} \notag \\
	&= \frac{1}{\ell} \cdot \E{w(M^{(\ell)})} \,.
	\end{align}
	Combining (\ref{eq:matchingWeightTentativeEdge}) and (\ref{eq:matchingWeightRoundL}) concludes the proof.
\end{proof}

However, we only gain the weight of the tentative edge $e^{(\ell)} = \{l,r\}$ if it can be added to the matching, i.e., if $r$ has not been matched previously. The next lemma bounds the probability for this event from below.

\begin{lemma}[\cite{DBLP:conf/esa/KesselheimRTV13}]
	\label{lemma:MatchingPrUnmatchedVertex}	
	Let $\xi(r,\ell)$ be the event that the offline vertex $r \in R$ is unmatched after round $\ell \geq cn+1$.
	It holds that $\Pr{\xi(r,\ell)} \geq \frac{cn}{\ell}$.
\end{lemma}
\begin{proof}
	In each round $k$, the vertex $r$ can only be matched if it is incident to the tentative edge 
	$e^{(k)} \in M^{(k)}$ of this round,
	i.e., $e^{(k)} = \{l,r\}$ where $l \in L$ is the online vertex of round $k$.
	As $l$ can be seen as uniformly drawn among all $k$ visible nodes
	(particularly, independent of the order of the previous $k-1$ items),
	$l$ has probability $1/k$ to arrive in round $k$. Consequently, $r$ is not matched in round $k$
	with probability $1- 1/k$.
	This argument applies to all rounds $cn+1,\ldots,\ell$. Therefore,
	\[ 
	\Pr{\xi(r,\ell)}
	\geq \prod_{k=cn+1}^{\ell} 1 - \frac{1}{k}
	= \prod_{k=cn+1}^{\ell} \frac{k-1}{k}
	= \frac{cn}{\ell} \,.
	\qedhere
	\]
\end{proof}

Using Lemmas~\ref{lemma:MatchingExpectedWeight} and \ref{lemma:MatchingPrUnmatchedVertex}, 
we can bound the competitive ratio of $\mathcal{A}_L$ in the following lemma.
Note that we obtain the optimal $(1/e)$-competitive algorithm from \cite{DBLP:conf/esa/KesselheimRTV13} for $c=1/e$ and $d=1$.

\begin{lemma}
	\label{lemma:MatchingCompetitiveRatio}
	It holds that $\E{\mathcal{A}_L} \geq \left( c \ln \frac{d}{c} - o(1) \right) \opt_L$.
\end{lemma}
\begin{proof}
	Let $A_\ell$ be the gain of the matching weight in round $\ell$.
	As the tentative edge $e^{(\ell)} = \{l,r\}$ can only be added if $r$ has not been matched in a previous round,
	we have $\E{A_\ell} = \E{w(e^{(\ell)})} \Pr{\xi(r,\ell)}$ for the event $\xi(r,\ell)$ from 
	Lemma~\ref{lemma:MatchingPrUnmatchedVertex}.
	Therefore, from Lemmas~\ref{lemma:MatchingExpectedWeight} and \ref{lemma:MatchingPrUnmatchedVertex}, we have
	$ \E{A_\ell} \geq \frac{1}{n} \opt_L \frac{cn}{\ell} = \frac{c}{\ell} \opt_L $.
	Summing over all rounds from $cn+1$ to $dn$ yields
	\[
	\E{\mathcal{A}_L}
	= \sum_{\ell=cn+1}^{dn} \E{A_\ell} 
	\geq \left( c \sum_{\ell=cn+1}^{dn}  \frac{1}{\ell} \right) \opt_L	
	\geq \left( c \ln \frac{d}{c} - \frac{1 - c/d}{n}\right) \opt_L \,.
	\]	
	The last inequality follows from 
	$\sum_{\ell=cn+1}^{dn} \frac{1}{\ell} = \left(\sum_{\ell=cn}^{dn-1} \frac{1}{\ell} \right) - \frac{1}{cn} + \frac{1}{dn}$ and, according to Fact~\ref{fact:approxSumByIntegralDec}, $\sum_{\ell=cn}^{dn-1} \frac{1}{\ell} 
	\geq \int_{cn}^{dn} \frac{1}{\ell} \intD{\ell} = \ln \frac{d}{c}$.
\end{proof}

\subsection{Small Options}

For small options, we use the LP-based algorithm from \cite[Sec.~3.3]{DBLP:journals/siamcomp/KesselheimRTV18} and analyze it within our algorithmic framework.
In order to make this paper self-contained, we give a linear program for fractional GAP (LP~1), 
the algorithm, and its corresponding proofs.
\begin{align*}
\text{maximize }  &\sum_{\substack{i \in I_S \\ r \in R}} v_{i,r} x_{i,r} \\
\text{subject to } &\sum_{i \in I_S} s_{i,r} x_{i,r} \leq W_r & \forall r \in R \\
&\sum_{r \in R} x_{i,r} \leq 1 & \forall i \in I_S \\
& 0 \leq x_{i,r} \leq 1 & \forall (i,r) \in I_S \times R
\tag{LP~1}
\end{align*}
\begin{algorithm}
	\SetKwInOut{Input}{Input}
	\SetKwInOut{Output}{Output}
	\Input{Random order sequence of $(1/2)$-small options, \\
		parameter $d \in (0,1)$.}
	\Output{Integral GAP assignment.}
	Let $\ell$ be the current round and $i$ be the online item of round $\ell$. \\
	\uIf{$1 \leq \ell \leq dn$} {Sampling phase -- do not assign any item.}
	\uIf{$dn+1 \leq \ell \leq n$} {
		Let $x^{(\ell)}$ be an optimal solution of LP~1 for $I_S(\ell)$. \\
		Choose a resource $r$ (possibly none), where $r$ has probability $x^{(\ell)}_{i,r}$. \\
		\uIf{the remaining capacity of $r$ is at least $(1/2)\cdot W_r$} {
			Assign $i$ to $r$.
		}
	}
	\caption{GAP algorithm for small options from \cite[Sec.~3.3]{DBLP:journals/siamcomp/KesselheimRTV18}.}
	\label{alg:lp}
\end{algorithm}

Let $\mathcal{A}_S$ be Algorithm~\ref{alg:lp}.
After a sampling phase of $dn$ rounds, in each round $\ell$, the algorithm computes 
an optimal solution $x^{(\ell)}$ of LP~1 for $I_S(\ell)$.
Here, $I_S(\ell)$ denotes the instance of small options revealed so far.
Now, the decision to which resource the current online item $i$ is assigned, if at all, 
is made at random using $x^{(\ell)}$:
Resource $r \in R$ is chosen with probability $x^{(\ell)}_{i,r}$ and the item stays unassigned with
probability $1 - \sum_{r \in R} x^{(\ell)}_{i,r}$. 
Note that the item can only be assigned to the chosen resource if its remaining capacity is at least $(1/2) \cdot W_r$.

To analyze Algorithm~\ref{alg:lp}, we consider the gain of profit in round $\ell \geq dn+1$, denoted
by $A_\ell$.
For this purpose, let $i^{(\ell)}$ be the item of that round and $r^{(\ell)}$ the resource chosen by the algorithm.
Now, it holds that 
$\E{A_\ell} = \E{v_{i^{(\ell)},r^{(\ell)}}} \cdot \Pr{\text{$i^{(\ell)}$ can be assigned to $r^{(\ell)}$}}$,
where in the first term, the expectation is over the item arriving in round $\ell$ and the resource chosen by
the algorithm. The latter term only depends on the resource consumption of $r^{(\ell)}$
in earlier rounds. In the next two lemmas, we give lower bounds for both terms.
As in the proofs of Section~\ref{sec:gapLarge},
it is helpful to construct the random permutation of the first $\ell$ items in two independent steps: First, the set of $\ell$ visible items is drawn uniformly, without determining the order of items. Second, the online item arriving in round $\ell$ is drawn uniformly from this set.

\begin{lemma}[{\cite[Sec.~3.3]{DBLP:journals/siamcomp/KesselheimRTV18}}]
	\label{lemma:LPexpectedProfit}
	For any round $\ell \geq dn+1$, we have 
	$\E{v_{i^{(\ell)},r^{(\ell)}}} \geq \frac{1}{n} \opt_S$.
\end{lemma}
\begin{proof}
	The proof is similar to the proof of Lemma~\ref{lemma:MatchingExpectedWeight}.
	As we consider a fixed round $\ell$, we write $i$ and $r$ instead of $i^{(\ell)}$ and $r^{(\ell)}$
	for ease of presentation.
	Further, we write $v(\alpha) := \sum_{j \in I_S} \sum_{s \in R} \alpha_{j,s} v_{j,s}$ for the
	profit of a fractional assignment $\alpha$.
	
	First, the set of visible items $Q$ in round $\ell$ is drawn uniformly at random among all subsets of $\ell$ items.
	Let $x^{(n)}$ be an optimal (offline) solution to LP~1 and
	let $x^{(n)}|_Q$ denote the restriction of $x^{(n)}$ to the items in $Q$, i.e.,
	$(x^{(n)}|_Q)_{j,s} = x^{(n)}_{j,s}$ if $j \in Q$ and $(x^{(n)}|_Q)_{j,s} = 0$ if $j \notin Q$.
	Since $x^{(n)}|_Q$ is a feasible and $x^{(\ell)}$ is an optimal solution for $Q$, we have
	$\E{v(x^{(\ell)})} \geq \E{v(x^{(n)}\mid_Q)}$.
	As each item has the same probability of $\ell/n$ to be in $Q$, it holds that
	\begin{equation}
	\label{eq:lpProfit1}
	\E{v(x^{(\ell)})} 
	\geq \E{v(x^{(n)}\mid_Q)}
	= \sum_{j \in I_S} \sum_{s \in R} \Pr{j \in Q} \cdot x^{(n)}_{j,s} \cdot v_{j,s}
	= \frac{\ell}{n} v(x^{(n)}) 
	\geq \frac{\ell}{n} \opt_S \,.	
	\end{equation}
	
	In the second step, the online item of round $\ell$ is determined by choosing one item from $Q$ uniformly at random.
	Let $\mathcal{X}$ be the domain of $x^{(\ell)}$ and $x' \in \mathcal{X}$. We have
	\begin{equation}
	\E{v_{i,r} \mid x^{(\ell)} = x'} 
	= \sum_{j \in Q} \sum_{s \in R} \Pr{j=i, s=r} v_{j,s} 
	= \sum_{j \in Q} \sum_{s \in R} \frac{1}{\ell} \cdot x'_{j,s} \cdot v_{j,s}
	= \frac{1}{\ell} v(x') \,,
	\end{equation}
	where we used that each item from $Q$ arrives in round $\ell$ with probability $1/\ell$ and the algorithm assigns item $j$ to resource $s$ with probability $x'_{j,s}$, given $x^{(\ell)} = x'$. By the law of total expectation,
	it follows that $\E{v_{i,r}} = \frac{1}{\ell} \E{v(x^{(\ell)})}$.
	Combining with (\ref{eq:lpProfit1}) gives the claim.
\end{proof}

Hence, by the previous lemma, the expected gain of profit in each round is at least a $(1/n)$-fraction of $\opt_S$,
supposing the remaining resource capacity is large enough.
The probability for the latter event is considered in the following lemma.
Here, a crucial property is that we deal with $\delta$-small options.
As in Section~\ref{sec:ksSmallAnalysis}, we define $\Delta=\frac{1}{1-\delta}$.

\begin{lemma}
	\label{lemma:LPsuccessfullPacking}
	For any round $\ell \geq dn+1$, it holds that
	\[\Pr{\text{$i^{(\ell)}$ can be assigned to $r^{(\ell)}$}} \geq \frac{c}{d} \left(1 - \Delta \ln \frac{\ell}{dn}\right) \,. \]
\end{lemma}
\begin{proof}
	Let $\xi$ be the event that no item is assigned to $r$ after round $dn$.
	Note that $\xi$ does not necessarily hold, since $\mathcal{A}_L$ might already have assigned
	items to $r$ in earlier rounds. By Lemma~\ref{lemma:MatchingPrUnmatchedVertex},	$\Pr{\xi} \geq \frac{c}{d}$.
	Therefore, it is sufficient to show 
	$\Pr{\text{$i^{(\ell)}$ can be assigned to $r^{(\ell)}$} \mid \xi} \geq 1 - \Delta \ln \frac{\ell}{dn}$.
	
	For this purpose, assume that $\xi$ holds and
	let $X$ denote the resource consumption of $r$ after round $\ell - 1$.
	Further, let $X_k$ be the resource consumption of $r$ in round $k < \ell$.
	We have $X = \sum_{k=dn+1}^{\ell-1} X_k$.
	Let $Q$ be the set of $k$ visible items in round $k$.
	The set $Q$ can be seen as uniformly drawn from all $k$-item subsets and any item $j \in Q$ is the current
	online item of round $k$ with probability $1/k$.
	Now, the algorithm assigns any item $j$ to resource $r$ with probability $x_{j,r}^{(k)}$, thus
	\begin{equation}
	\label{eq:LPexpectedReousrceConsumptionRoundK}
	\E{X_k}
	= \sum_{j \in Q} \Pr{j \text{ occurs in round } k} s_{j,r} x_{j,r}^{(k)}
	= \frac{1}{k} \sum_{j \in Q} s_{j,r} x_{j,r}^{(k)}
	\leq \frac{W_r}{k} \,,
	\end{equation}
	where the last inequality follows from the capacity constraint for resource $r$ in LP~1.
	By linearity of expectation and inequality~(\ref{eq:LPexpectedReousrceConsumptionRoundK}), 
	the expected resource consumption up to round $\ell$ is thus
	\begin{equation}
	\label{eq:LPexpectedTotalResourceConsumption}
	\E{X} = \sum_{k=dn+1}^{\ell-1} \E{X_k} \leq \sum_{k=dn+1}^{\ell-1} \frac{W_r}{k} \leq W_r \ln \frac{\ell}{dn} \,.
	\end{equation}
	Now, since $i^{(\ell)}$ is $\delta$-small, $X < (1-\delta) W_r$ implies $X + s_{i^{(\ell)},r^{(\ell)}} \leq W_r$,
	in which case the assignment is feasible.
	Using (\ref{eq:LPexpectedTotalResourceConsumption}) and Markov's inequality, we obtain
	\begin{equation*}
	\Pr{X < (1-\delta) W_r }
	= 1 - \Pr{X \geq (1-\delta) W_r}
	\geq 1 - \frac{\E{X}}{(1-\delta) W_r} 
	\geq 1 - \Delta \ln \frac{\ell}{dn} \,.
	\qedhere
	\end{equation*}
\end{proof}
The next lemma finally gives the competitive ratio of $\mathcal{A}_S$.
\begin{lemma}
	\label{lemma:LPcompetitiveRatio}
	It holds that
	\[
	\E{\mathcal{A}_S} \geq \frac{c}{d} \left( (1-d) (1+\Delta)  - \Delta \cdot \left( 1 + \frac{1}{n} \right) \cdot \ln \frac{1}{d} \right) \opt_S \,.
	\]
\end{lemma}
\begin{proof}
	We add the expected profits in single rounds using Lemmas~\ref{lemma:LPexpectedProfit} and~\ref{lemma:LPsuccessfullPacking}.
	\begin{align*}
	\E{\mathcal{A}_S} 
	&= \sum_{\ell=dn+1}^{n} \E{A_\ell} \\
	&= \sum_{\ell=dn+1}^{n}  \E{v_{i^{(\ell)},r^{(\ell)}}} \Pr{\text{$i^{(\ell)}$ can be assigned to $r^{(\ell)}$}} \\
	&\geq \sum_{\ell=dn+1}^{n}  \frac{1}{n} \opt_S \frac{c}{d} \left(1 - \Delta \ln \frac{\ell}{dn}\right) \\
	&= \frac{c}{dn} \left( \sum_{\ell=dn+1}^{n}  1 - \Delta \ln \frac{\ell}{dn} \right) \opt_S \\
	&= \frac{c}{dn} \left( n - dn - \Delta \sum_{\ell=dn+1}^{n}  \ln \frac{\ell}{dn} \right) \opt_S \,.
	\end{align*}
	Since $\frac{\ell}{dn}$ is monotonically increasing in $\ell$, we have
	\[
	\sum_{\ell=dn+1}^{n} \ln \frac{\ell}{dn} 
	= \left(\sum_{\ell=dn}^{n-1} \ln \frac{\ell}{dn}\right) + \ln \frac{n}{nd}
	\leq \left( \int_{dn}^{n} \ln \frac{\ell}{dn} \intD{\ell}\right) + \ln \frac{1}{d}
	\]
	by Fact~\ref{fact:approxSumByIntegralInc}.
	The integral $\int_{dn}^{n} \ln \frac{\ell}{dn} \intD{\ell}$ evaluates to $n \cdot \left( \ln \frac{1}{d} - 1 + d \right)$, so combining the previous inequalities yields
	\begin{align*}
	\E{\mathcal{A}_S} 
	&> \frac{c}{d} \left( 1 - d - \Delta \cdot \left( \ln \frac{1}{d} - 1 + d \right) - \frac{\Delta}{n} \cdot \ln \frac{1}{d}\right) \opt_S \\
	&= \frac{c}{d} \left( (1 - d) (1+\Delta) - \Delta \cdot \left( 1 + \frac{1}{n} \right) \cdot \ln \frac{1}{d} \right) \opt_S \,.
	\qedhere
	\end{align*}
\end{proof}
Note that we obtain the same competitive ratio as in Lemma~\ref{lemma:KSsmallItemsFinal}.

\subsubsection{Remark}
The setting of large capacities (compared to the respective resource demands) has been addressed in
several papers \cite{DBLP:conf/approx/AlaeiHL13,DBLP:conf/wine/FeldmanKMMP09,DBLP:conf/wine/ZhouCL08}.
For instance, such settings arise in online auctions, where the budgets are very high compared to single bids.
Although the algorithm $\mathcal{A}_S$ is not tailored for this setting,
a corresponding bound can be obtained easily from Lemma~\ref{lemma:LPcompetitiveRatio}.
Setting $c=d$ clearly maximizes the performance of $\mathcal{A}_S$ with respect to $\opt_S$,
thus the factor $c/d$ vanishes.
Assuming that the maximum resource demand is $\delta \to 0$, 
the competitive ratio of $\mathcal{A}_S$ tends to $2 (1-d) - \ln \frac{1}{d}$, since $\Delta \to 1$. 
This function is maximized for $d=1/2$, yielding a competitive ratio of $1 - \ln (2) \geq 0.3068$.

\subsection{Proof of Theorem~\ref{thm:GAPmainTheorem}}

Finally, we prove our main theorem for GAP.

\begin{proof}[Proof of Theorem~\ref{thm:GAPmainTheorem}]
	We set the threshold between large and small options to $\delta = 1/2$ and consider Algorithm~\ref{alg:sequentialAlg} with the algorithms $\mathcal{A}_L$ and $\mathcal{A}_S$ as defined previously.
	By Lemma~\ref{lemma:MatchingCompetitiveRatio}, the expected gain of profit in rounds $cn+1,\ldots,dn$ is
	$\E{\mathcal{A}_L} \geq \left( c \ln \frac{d}{c} - o(1) \right) \opt_L$.
	In the following rounds, we gain
	\[\E{\mathcal{A}_S} \geq \frac{c}{d} \left(3 (1-d) - 2 \ln \frac{1}{d} - o(1) \right) \opt_S \]
	according to Lemma~\ref{lemma:LPcompetitiveRatio} (with $\Delta = 2$).
	Setting
	$c=0.5261$ and $d=0.6906$ gives
	$c \ln \frac{d}{c} \approx \frac{c}{d} \left(3 (1-d) - 2 \ln \frac{1}{d}\right)$
	and thus, using $\opt_L + \opt_S \geq \opt$,
	\begin{align*}
	\E{\mathcal{A}_L} + \E{\mathcal{A}_S} 
	&\geq \frac{c}{d} \left(3  (1-d) - 2 \ln \frac{1}{d} - o(1) \right) \left(\opt_L + \opt_S \right) \\
	&\geq \left( \frac{1}{6.99} - o(1) \right)\opt \,.
	\qedhere
	\end{align*}
\end{proof}

\paragraph{Acknowledgements.}
We thank the anonymous reviewers for many valuable comments on an earlier version of this manuscript.

\bibliographystyle{plain}      
\bibliography{bibliography}   

\end{document}